\documentclass[prl,twocolumn,
preprintnumbers,superscriptaddress,amsmath,amssymb]{revtex4}

\usepackage{graphicx,braket}
\usepackage{subfigure}
\usepackage{mathrsfs}
\usepackage{amsfonts}
\usepackage{amsmath,comment}
\usepackage[colorlinks,linkcolor=blue,citecolor=blue]{hyperref}

\begin{document}

\title{Discrete Time-Crystalline Order in Cavity and Circuit QED Systems}
\author{Zongping Gong}
\affiliation{Department of Physics, University of Tokyo, 7-3-1 Hongo, Bunkyo-ku, Tokyo 113-0033, Japan}
\author{Ryusuke Hamazaki}
\affiliation{Department of Physics, University of Tokyo, 7-3-1 Hongo, Bunkyo-ku, Tokyo 113-0033, Japan}
\author{Masahito Ueda}
\affiliation{Department of Physics, University of Tokyo, 7-3-1 Hongo, Bunkyo-ku, Tokyo 113-0033, Japan}
\affiliation{RIKEN Center for Emergent Matter Science (CEMS), Wako, Saitama 351-0198, Japan}
\date{\today}

\begin{abstract}
Discrete time crystals are a recently proposed and experimentally observed out-of-equilibrium dynamical phase of Floquet systems, where the stroboscopic evolution of a local observable repeats itself at an integer multiple of the driving period. We address this issue in a driven-dissipative setup, focusing on the modulated open Dicke model, which can be implemented by cavity or circuit QED systems. In the thermodynamic limit, we employ semiclassical approaches and find rich dynamical phases on top of the discrete time-crystalline order. In a deep quantum regime with few qubits, we find clear signatures of a transient discrete time-crystalline behavior, which is absent in the isolated counterpart. We establish a phenomenology of dissipative discrete time crystals by generalizing the Landau theory of phase transitions to Floquet open systems.
\end{abstract}

\maketitle

\emph{Introduction.---}
Phases and phase transitions of matter are key concepts for understanding complex many-body physics \cite{Chaikin2000,Sachdev2011}. Recent experimental developments in various quantum simulators, such as ultracold atoms \cite{Bloch2008,Bloch2012}, trapped ions \cite{Wineland2003,Blatt2012} and superconducting qubits \cite{Koch2012,Devoret2013}, motivate us to seek for quantum many-body systems out of equilibrium \cite{Polkovnikov2011,Eisert2015,Moessner2017}, 
such as many-body localized phases \cite{Pal2010,Altman2015,Nandkishore2015,Abanin2015,Moessner2015,Choi2016} and Floquet topological phases \cite{Kitagawa2010,Jiang2011,Lindner2013,Lindner2016,Keyserlingk2016a,Else2016,Potter2016,Vishwanath2016}.

In recent years, much effort has been devoted to periodically driven (Floquet) quantum many-body systems that break the discrete time-translation symmetry (TTS) \cite{Sacha2017}.
In contrast to the continuous TTS breaking \cite{Wilczek2012,Wilczek2012b,Duan2012} that has turned out to be impossible at thermal equilibrium \cite{Bruno2013,Watanabe2015}, the discrete TTS breaking has been theoretically proposed \cite{Sacha2015,Khemani2016,Nayak2016,Keyserlingk2016b,Yao2017} and experimentally demonstrated \cite{Zhang2017,Choi2017}. Phases with broken discrete TTS feature discrete time-crystalline (DTC) order characterized by periodic oscillations of physical observables with period $nT$, where $T$ is the Floquet period and $n=2,3,\cdots$. The DTC order is expected to be stabilized by many-body interactions against variations of driving parameters. Note that the system is assumed to be in a localized phase \cite{Khemani2016,Nayak2016,Yao2017,Zhang2017} or to have long-range interactions \cite{Choi2017,Ho2017,Russomanno2017}. Otherwise, the DTC order only exists in a prethermalized regime \cite{Nayak2017,Sheng2017} since the system will eventually be heated to a featureless infinite-temperature state due to persistent driving \cite{DAlessio2014,Moessner2014,Ikeda2014}.

While remarkable progresses are being made concerning the DTC phase, most studies focus on isolated systems. Indeed, as has been experimentally observed \cite{Zhang2017,Choi2017} and theoretically investigated \cite{Moessner2017b}, the DTC order in an open system is usually destroyed by decoherence. On the other hand, it is known that dissipation and decoherence can also serve as resources for quantum tasks such as quantum computation \cite{Verstraete2009} and metrology \cite{delCampo2017}. From this perspective, it is natural to ask whether the DTC order exists and can even be stabilized in open systems \cite{Breuer2002}. Such a possibility has actually been pointed out in Ref.~\cite{Nayak2017}, but neither a detailed theoretical model nor a concrete experimental implementation is presented.  
 
\begin{figure}
\begin{center}
        \includegraphics[width=8.5cm, clip]{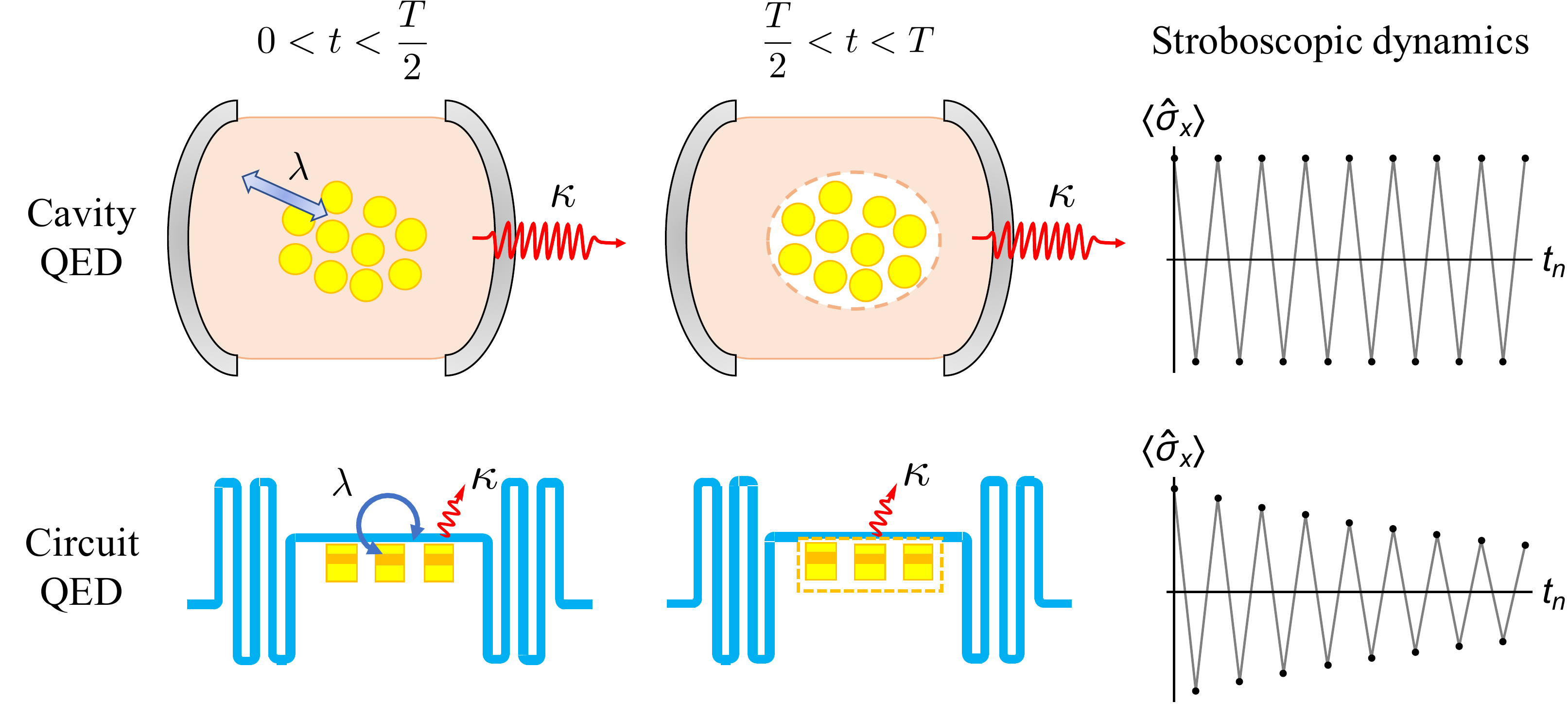}
      \end{center}
   \caption{(color online). Cavity and circuit QED setups for realizing the DTC order. In the first (second) half of a Floquet period $T$, we switch on (off) the coupling $\lambda$ between light and (artificial) atoms. For sufficiently large $\lambda$, almost persistent DTC order in the stroboscopic dynamics of a local observable is expected for an ensemble of a large number of atoms in an optical cavity, while transient DTC behavior can be observed for few superconducting qubits coupled to a microwave transmission line. Here $\kappa$ denotes the loss rate of (microwave) photons.}
   \label{fig1}
\end{figure}

\begin{figure*}
\begin{center}
       \includegraphics[width=16cm, clip]{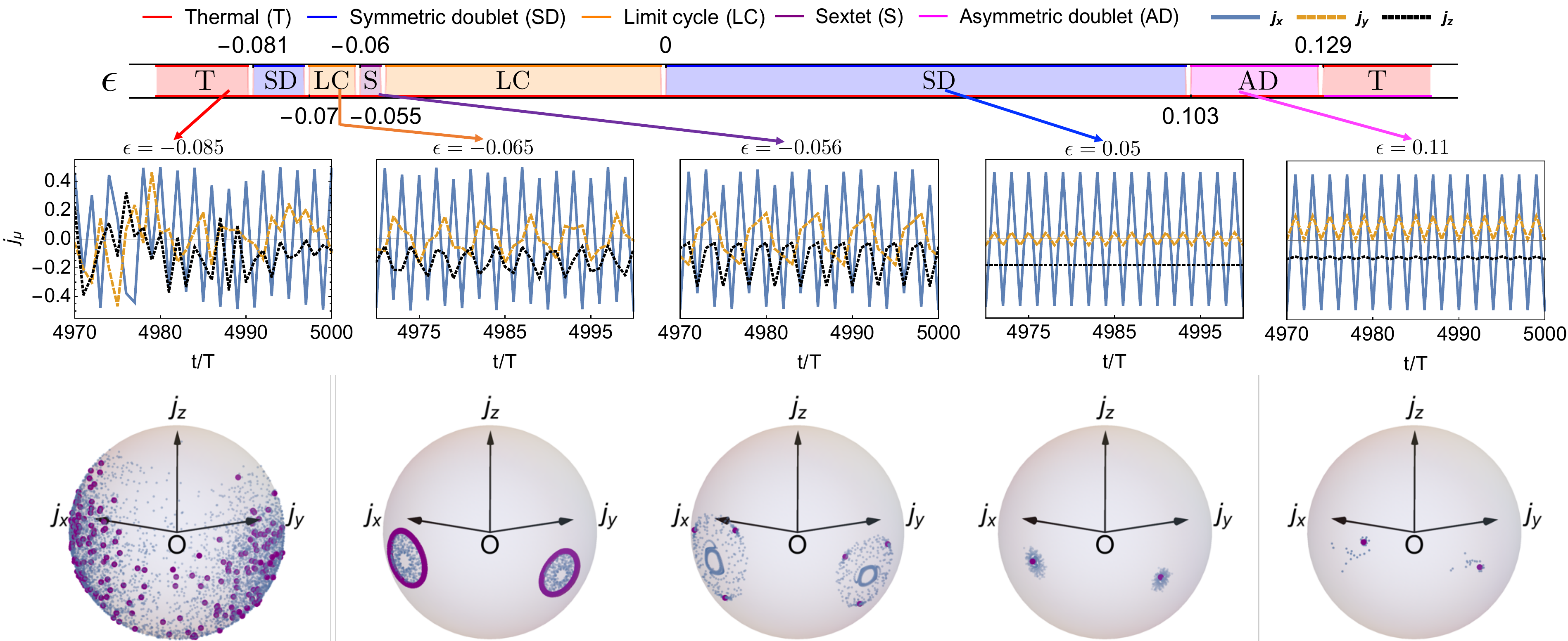}
       \end{center}
   \caption{(color online). Dynamical phase diagram (top), typical stroboscopic dynamics (middle), and trajectories (bottom) of the atomic pseudospin for atom-light coupling $\lambda=1$ and photon-loss rate $\kappa=0.05$. Top: As the detuning $\epsilon$ [see Eq. (\ref{detuning_frequency})] is varied, five different dynamical phases emerge: thermal (T, red), symmetric period doubling (normal DTC order, SD, blue), limit-cycle pair (LC, orange), period sextupling (S, purple), and asymmetric period doubling (AD, magenta). The phase boundaries are marked in white with resolution $10^{-3}$. Middle: Typical stroboscopic dynamics of $j_\mu\equiv\frac{1}{N}\langle\hat J_\mu\rangle$ ($\mu=x$ (solid blue), $y$ (dashed orange), $z$ (dotted black)) for the last $30$ periods of the entire $5000$-period evolution. Bottom: Full stroboscopic phase-space-point trajectories (light blue) and those of the last $200$ periods (purple) projected on the pseudospin Bloch sphere.}
   \label{fig2}
\end{figure*}

In this Letter, we propose a concrete open-system setup for realizing the DTC order by using a prototypical dissipative model --- a modified open Dicke model \cite{Dicke1954,Zakowicz1975,Brandes2003}, which describes a collective light-atom interaction in the presence of interaction modulation and photon loss. This model is relevant to cavity QED systems based on cold atoms \cite{Esslinger2010,Esslinger2011,Esslinger2013,Parkins2014} and circuit QED systems based on superconducting qubits \cite{Blais2004,Ciuti2010,Marquardt2011,Wallraff2014,Rabl2016,Nakamura2016,Semba2017}. As schematically illustrated in Fig.~\ref{fig1}, the DTC order manifests itself through periodic switch-on and switch-off of a sufficiently strong light-atom coupling. For the cavity QED case, we consider the thermodynamic limit and find unexpectedly rich dynamical phases as the detuning parameter is varied 
(see Fig.~\ref{fig2}). For the circuit QED case, we examine a deep quantum regime with few qubits to find a clear transient DTC behavior even for two qubits, a minimal setup of superradiance \cite{Wallraff2014}. We also discuss a phenomenological model which demonstrates the exponentially long lifetime of the DTC order. These predictions should be testable in light of the state-of-the-art experimental developments in atomic, molecular and optical physics.

\emph{Modulated open Dicke model.---}  We consider $N$ identical two-level atoms in a single-mode cavity. Neglecting the atomic motional degrees of freedom, the dynamics of the system can be described by the open Dicke model \cite{Carmichael2007}:
\begin{equation}
\begin{split}
\frac{d{\hat\rho}_t}{dt}=\mathcal{L}(\lambda)\hat\rho_t=-i[\hat H(\lambda),\hat\rho_t]+\kappa\mathcal{D}[\hat a]\hat\rho_t,\\
\hat H(\lambda)=\omega\hat a^\dag\hat a+\omega_0\hat J_z+\frac{2\lambda}{\sqrt{N}}(\hat a+\hat a^\dag)\hat J_x,
\end{split}
\label{ODME}
\end{equation}
where $\mathcal{D}[\hat a]\hat\rho\equiv\hat a\hat\rho\hat a^\dag-\frac{1}{2}\{\hat a^\dag\hat a,\hat\rho\}$, $\hat a$ is the annihilation operator of the photon field, $\hat J_\mu\equiv\frac{1}{2}\sum^N_{j=1}\hat\sigma^\mu_j$ ($\mu=x,y,z$) is the collective atomic pesudospin operator, $\omega$, $\omega_0$, $\lambda$ and $\kappa$ are the optical frequency, the atomic frequency, the coupling strength and the photon-loss rate, respectively. 

It is known that, in the thermodynamic limit and when $\lambda$ exceeds $\lambda_{\rm c}=\frac{1}{2}\sqrt{\frac{\omega_0}{\omega}(\omega^2+\frac{\kappa^2}{4})}$, the open Dicke model exhibits a phase transition that breaks the $\mathbb{Z}_2$ symmetry characterized by the parity operator $\hat P\equiv e^{i\pi(\hat a^\dag\hat a+\hat J_z+\frac{N}{2})}$ \cite{Brandes2003,Esslinger2011}. For $\lambda>\lambda_{\rm c}$, we can construct an exact period-doubling Floquet dynamics as follows: Starting from one of the symmetry-broken steady states $\hat\rho_{\rm ss}$, in the first-half period, the dynamics is governed by Eq.~(\ref{ODME}), so $\hat\rho_{\rm ss}$ stays unchanged by definition. In the second-half period, we perform the parity operation on the system, so that the other steady state $\hat\rho'_{\rm ss}=\hat P\hat\rho_{\rm ss}\hat P$ is obtained at the end of the Floquet period. If we observe the system stroboscopically at $t_n=nT$, we should find $\hat\rho_{\rm ss}$ ($\hat\rho'_{\rm ss}$) for even (odd) $n$.

If the period doubling is robust against imperfection such as the deviation of the evolution in the second-half period from the parity operation, we can identify it as a DTC order. A straightforward way to introduce such imperfection is to switch off the atom-light coupling in the second-half period. That is, we modulate $\lambda$ in Eq.~(\ref{ODME}) periodically as  
\begin{equation}
\lambda_{t+T}=\lambda_t=\left\{\begin{array}{ll}
\lambda & 0\le t<\frac{T}{2};\\
0 & \frac{T}{2}\le t<T.
\end{array}\right.
\label{protocol}
\end{equation}
In the resonant ($\omega=\omega_0=\omega_T\equiv\frac{2\pi}{T}$) and isolated ($\kappa=0$) case, the state evolution during the second half of the period generates the parity operator up to an unimportant global phase, i.e., $\hat P=e^{-i\frac{T}{2}\hat H(0)+\frac{i\pi}{2}N}$. If we introduce a detuning between $\omega$ and $\omega_0$ as
\begin{equation}
\omega=(1-\epsilon)\omega_T,\;\;\;\;\omega_0=(1+\epsilon)\omega_T,
\label{detuning_frequency}
\end{equation}
we can control the degree of imperfection by $\epsilon$. Note that there is always a \emph{nonunitary} imperfection due to photon loss even for $\epsilon=0$. For simplicity, we set $\omega_T=1$ in the following discussion.

\emph{Dynamical phases 
in the thermodynamic limit.---} In the thermodynamic limit $N\to\infty$, the relative fluctuation in a local observable becomes negligible and the semiclassical approach is justified \cite{Brandes2012,Keeling2012,Zilberberg2015}. In terms of the scaled variables $x\equiv\frac{\langle\hat a+\hat a^\dag\rangle}{\sqrt{2N\omega}}$, $p\equiv\frac{i\langle\hat a^\dag-\hat a\rangle}{\sqrt{2N/\omega}}$ and $\boldsymbol{j}\equiv(j_x,j_y,j_z)$ with $j_\mu\equiv\frac{1}{N}\langle\hat J_\mu\rangle$ ($\mu=x,y,z$), the semiclassical dynamics governed by Eq.~(\ref{ODME}) reads \cite{SM}
\begin{equation}
\begin{split}
&\;\;\;\;\;\;\;\;\;\frac{d\boldsymbol{j}}{dt}=(\omega_0\boldsymbol{e}_z+2\lambda_t\sqrt{2\omega}x\boldsymbol{e}_x)\times\boldsymbol{j},\\
\frac{dx}{dt}&=p-\frac{\kappa}{2}x,\;\;\;\;\frac{dp}{dt}=-\omega^2 x-\frac{\kappa}{2}p-2\lambda_t\sqrt{2\omega}j_x.
\end{split}
\label{semicl}
\end{equation}
Note that the $\mathbb{Z}_2$ symmetry is maintained, since Eq.~(\ref{semicl}) is invariant under the simultaneous sign reversal of $x$, $p$, $j_x$ and $j_y$. The dissipative phase transition \cite{Kessler2012} in the open Dicke model now becomes a dynamical phase transition known as the pitchfork bifurcation \cite{Keeling2012}, where the original unique attractor with $x_0=p_0=j_{x0}=j_{y0}=0$ and $j_{z0}=\frac{1}{2}$ becomes unstable and two new stable attractors with
$(j_{x\pm},j_{y\pm},j_{z\pm})=\frac{1}{2}(\pm\sqrt{1-\mu^2},0,-\mu)$ and
$(x_\pm,p_\pm)=\mp\frac{\sqrt{2\omega(1-\mu^2)}}{\omega^2+\kappa^2/4}(\lambda,\frac{\kappa}{2})$
 ($\mu\equiv\frac{\lambda^2_{\rm c}}{\lambda^2}$) emerge as the classical reductions from $\hat\rho_{\rm ss}$ and $\hat\rho'_{\rm ss}$. To be specific, we fix $\lambda=1$ and $\kappa=0.05$ in the following calculations and choose the initial state to be the ``$+$" attractor.

We solve the nonlinear differential equation (\ref{semicl}) up to $5000$ periods by using the Runge-Kutta method for different $\epsilon$ and map out the full dynamical phase diagram in the top row of Fig.~\ref{fig2} \cite{RS}. 
We find the normal DTC phase and the thermal phase, where the former respects the $\mathbb{Z}_2$ symmetry in which $j_x$, $j_y$, $x$, $p$ reverse their signs after one period, and the latter shows irregular trajectories that cover some areas of the pseudospin sphere (or in the quadrature ($x$-$p$) plane). Furthermore, we find symmetric limit-cycle pairs, where the steady orbit forms two closed loops in the phase space, period sextupling, and asymmetric period doubling, with $j_x$, $j_y$, $x$, $p$ taking on two different values that are not symmetric against inversion. In fact, we find even richer dynamical phases for other $\kappa$, such as higher-order period multipling and asymmetric limit-cycle pairs \cite{SM}. These phases can unambiguously be diagnosed by a measure of synchronization \cite{Tony2013,Lesanovsky2015,Nunnenkamp2017} and can systematically be understood by employing bifurcation theory \cite{May1976,Holmes1983,Kuznetsov1998,Xiao2007,Flicker2017a,Flicker2017b}.

We note that the dynamics of a generalized \emph{time-independent} open Dicke model, which has an additional Stark-shift term $\frac{U}{N}\hat J_z\hat a^\dag\hat a$ in $H(\lambda)$ in Eq.~(\ref{ODME}), has thoroughly been studied in Ref.~\cite{Keeling2012} based on the semiclassical analysis. While there are only single- (normal) and double-attractor (superradiant) phases for $U=0$, limit-cycle and multiple-attractor phases emerge for $U\neq0$. In contrast, in this Letter, the richness of dynamical phases arises from the \emph{time dependence} of $\lambda$ with $U=0$. Another distinction is that in Ref.~\cite{Keeling2012} the steady state picks up one of the attractors or the unique limit cycle, whereas in the present Letter the steady state goes around different fixed points or limit cycles in a stroboscopic manner.

\emph{Transient DTC behavior in the deep quantum regime.---} Let us move to the few-atom regime ($N\sim O(1)$) which is the case for circuit QED systems. We consider the modulated open Dicke model with $N=2$. We demonstrate that the interplay between strong coupling and dissipation causes a DTC behavior for unexpectedly long periods even in this deep quantum regime. By unexpectedly long we mean that the DTC transient lasts much longer than the decay time $\kappa^{-1}\sim3T$.

\begin{figure}
\begin{center}
\includegraphics[width=8.6cm]{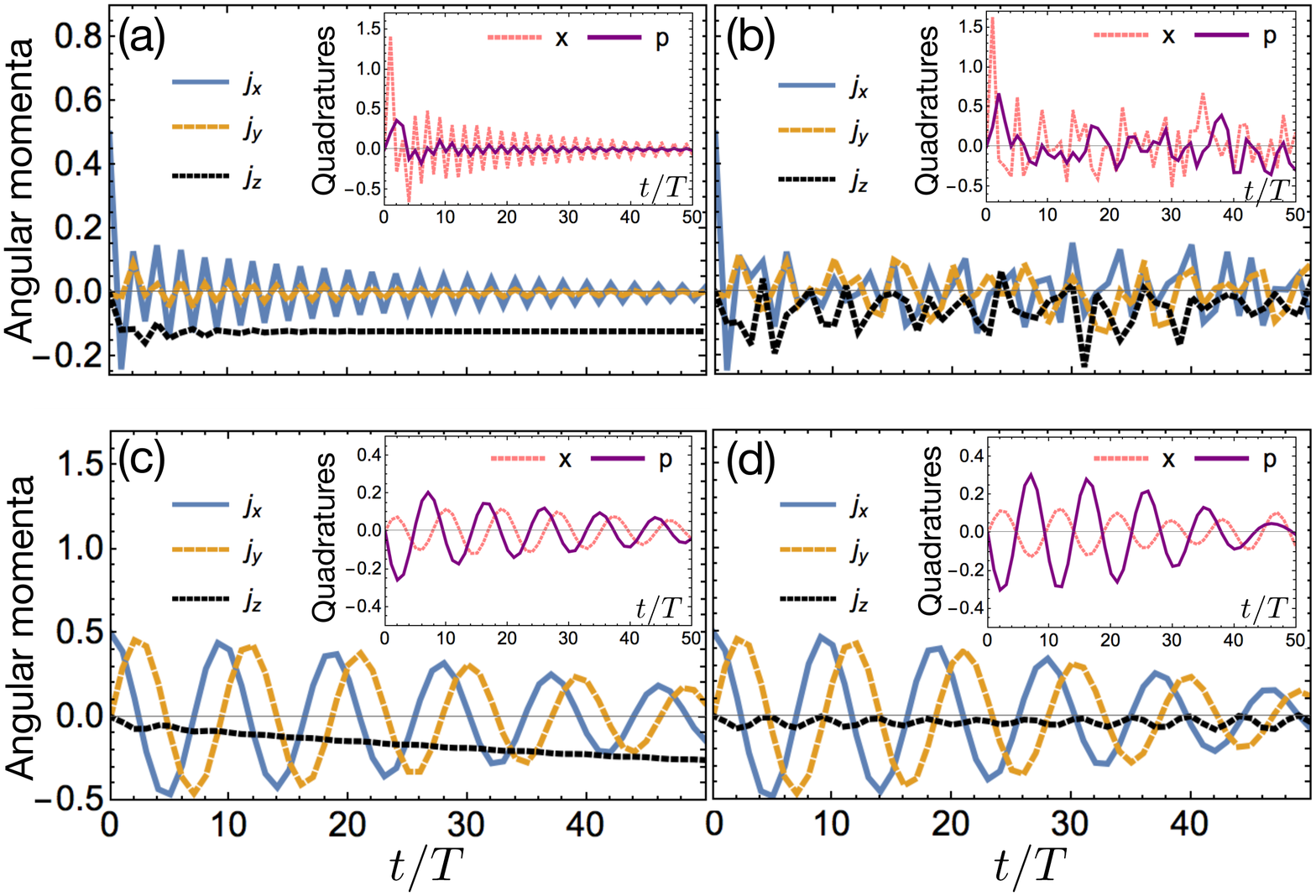}
\caption{(color online).
(a) Stroboscopic dissipative dynamics of the scaled angular momenta of $j_x$ (solid), $j_y$ (dashed), and $j_z$ (dotted) in the two-qubit Dicke model with $\kappa=0.05,\epsilon=0.1$ and $\lambda=1$ (strong coupling). The inset shows quadratures, $x$ (dotted) and $p$ (solid). (b) Stroboscopic dynamics for isolated systems ($\kappa=0,\epsilon=0.1$, and $\lambda=1$) in the strong-coupling regime. (c) Stroboscopic dissipative dynamics in the weak-coupling regime ($\kappa=0.05,\epsilon=0.1$, and $\lambda=0.1$). (d) Stroboscopic unitary dynamics in the weak-coupling regime ($\kappa=0,\epsilon=0.1$, and $\lambda=0.1$). Only (a) shows a DTC transient.
The initial state is always $\ket{\Rightarrow}\otimes\ket{0}$, where $\ket{\Rightarrow}$ is the eigenstate of $\hat{J}_x$ with eigenvalue $N/2\:(N=2)$ and $\ket{0}$ is the photon vacuum.
}
\label{deepq}
\end{center}
\end{figure}

We employ the exact diagonalization approach to solving the Floquet-Lindblad dynamics governed by Eqs.~(\ref{ODME}) and (\ref{protocol}) under a truncation up to 16 photons. Figure \ref{deepq} (a) shows the obtained stroboscopic dynamics of the scaled angular momenta $j_\mu$ and quadratures $x,p$ (inset) in the strong-coupling regime, where $\kappa=0.05,\epsilon=0.1$ and $\lambda=1$. The initial state is chosen to be $\ket{\Rightarrow}\otimes\ket{0}$, where $\ket{\Rightarrow}\equiv\bigotimes^N_{j=1}\ket{\rightarrow}$ is the eigenstate of $\hat{J}_x$ with eigenvalue $N/2\:(N=2)$ and $\ket{0}$ is the photon vacuum. We clearly see that $j_x$ and $x$ start oscillating with a period of $2T$ after $t\sim 5T$, which persists even at $t\sim 50T$. This result shows that our strong-coupling modulated open Dicke model features a DTC transient even in the deep quantum regime before reaching the stationary state. For the sake of comparison, we show in Fig.~\ref{deepq} (b) the stroboscopic dynamics for an isolated Dicke model ($N=2,\kappa=0,\epsilon=0.1,\lambda=1$) starting from the same initial state. We can see that the expectation value of each observable randomly fluctuates and does not have temporal order in contrast to its dissipative counterpart.

We note that no DTC transient emerges in the weak-coupling regime.
Figure \ref{deepq} (c) shows the Floquet dynamics for an open ($\kappa=0.05$) Dicke model with $\epsilon=0.1$ and $\lambda=0.1$. The low-frequency oscillation has a period around $T/\epsilon$ which is susceptible to detuning $\epsilon$. This is similar to the observation that the DTC order is fragile in noninteracting spin systems \cite{Yao2017,Zhang2017}. A similar dynamics is found in a weakly coupled isolated Dicke system ($\kappa=0,\epsilon=0.1$ and $\lambda=0.1$) as shown in Fig. \ref{deepq} (d). Thus, neither photon loss nor strong coupling alone gives rise to the DTC transient.

\begin{figure}
\begin{center}
        \includegraphics[width=8.6cm, clip]{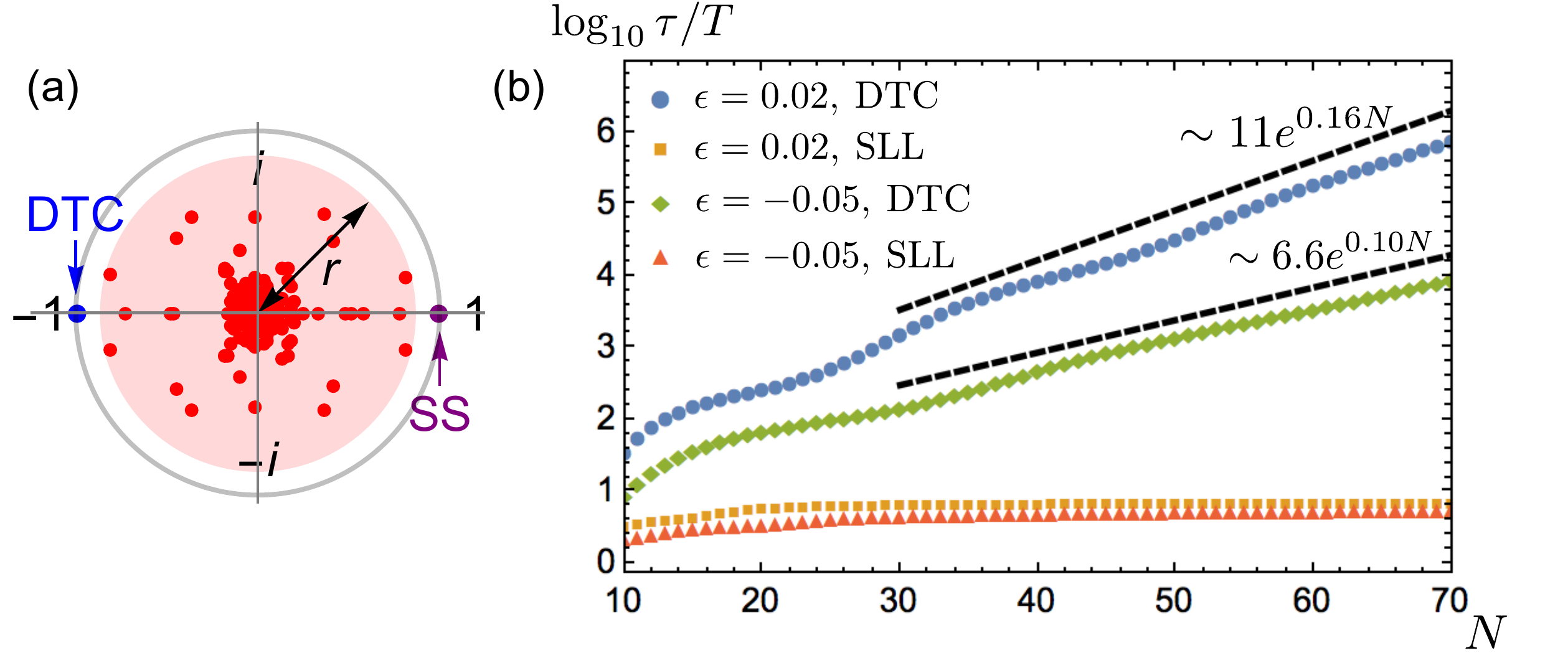}
      \end{center}
   \caption{(color online). (a) Typical Floquet-Lindblad spectrum of an open-system DTC. The DTC mode and the steady state (SS) locate at $-1+\delta$ and $1$, respectively, with $\delta\sim O(e^{-cN})$. The other modes locate in a disk (shaded) with radius $r<1$ for $\forall N$, so their lifetime is bounded by a constant $-\frac{T}{\ln r}$. (b) Finite-size scaling for the lifetime $\tau=-\frac{T}{\ln (1-\delta)}$ of the DTC and the second longest-lived (SLL) modes in the Floquet-Lindblad-Landau model (\ref{GFLLG}) for $\epsilon=0.02$ and $-0.05$.}
      \label{fig4}
\end{figure}

\emph{Floquet-Lindblad-Landau theory.---} With all the obtained numerical results in mind, we now establish a general phenomenology for such open-system DTC. As illustrated in Fig.~\ref{fig4} (a), the eigenvalues of the Floquet-Lindblad superoperator $\mathcal{U}_{\rm F}\equiv\mathcal{T}e^{\int^T_0dt\mathcal{L}(\lambda_t)}$ generally locate inside the unit circle in the complex plane, except for the steady state which always locates at $1$. Even if the initial state is a complex mixture of many eigenmodes, the state will eventually be described by fewer modes due to an exponential decay during time evolution. A semiclassical picture of this process is the convergence to attractors. When the state is described as a mixture of two eigenmodes, it can exhibit oscillatory DTC behavior with the double period if the distinguishably long-lived mode other than the steady state has a negative eigenvalue close to $-1$ \cite{SM}.

An important question is how the lifetime of this DTC mode scales with $N$. A natural expectation is that it becomes \emph{exponentially} long with increasing $N$, since the underlying dissipative phase transition features an exponentially small damping gap \cite{Ciuti2017}. However, it is highly nontrivial to find whether this is the case even in a Floquet open system. It turns out to be difficult to handle this problem numerically in the modulated open Dicke model. This difficulty emphasizes the importance of scalable circuit-QED-based quantum simulation with up to tens of qubits \cite{Koch2017}. Nevertheless, we can gain qualitative insights by considering a numerically tractable effective theory for the photon field:
\begin{equation}
\begin{split}
\frac{d\hat\rho_t}{dt}=&-i[\hat H_{\rm L}(\Omega_2(t),\Omega_4(t)),\hat\rho_t]+\kappa\mathcal{D}[\hat a]\hat\rho_t,\\
\hat H_{\rm L}(\Omega_2,\Omega_4)&=\omega\hat a^\dag\hat a-\frac{\Omega_2}{4}(\hat a^\dag+\hat a)^2+\frac{\Omega_4}{32N}(\hat a^\dag+\hat a)^4.
\end{split}
\label{GFLLG}
\end{equation}
These equations can be derived from the open Dicke model (\ref{ODME}) by adiabatically eliminating the atomic degrees of freedom under specific conditions \cite{SM}. Remarkably, Eq.~(\ref{GFLLG}) can be regarded as the Floquet-Lindblad generalization of the scalar-field Landau theory in $0+1$ dimension, and it is thus expected to capture the general qualitative features of a wide class of Floquet open systems in addition to the Dicke model. In Fig.~\ref{fig4} (b), we show the lifetime of the DTC (longest-lived) and that of the second longest-lived mode (except for the steady state) for a specific protocol $\Omega_4(t)=\Omega_2(t)=\Omega_2(t+T)$, where $\Omega_2(t)=1.5\omega$, $0\le t<\frac{\pi}{\omega}$ and $\Omega_2(t)=0$, $\frac{\pi}{\omega}\le t<T=(2-\epsilon)\frac{\pi}{\omega}$ and $\kappa=0.05\omega$. We do find an exponential scaling of the lifetime of the DTC order with respect to $N$ and the saturation of the lifetime of the second longest-lived mode. Note that the lifetime of a one-dimensional many-body localized DTC obeys the same exponential scaling in the system size \cite{Keyserlingk2016}, although the mechanism of DTC order is different \cite{Khemani2016,Keyserlingk2016b,Nayak2016,Yao2017}.

\emph{Summary and outlook.---} We have proposed a simple scheme for realizing DTC order in cavity and circuit QED systems via switching on and off of the atom-light coupling. In particular, we focus on the modulated open Dicke model both in the thermodynamic limit and in the deep quantum regime. In the former case, we find rich dynamical phases. In the latter case, we show that the interplay between dissipation and strong coupling gives rise to a clear transient DTC behavior. We demonstrate an exponentially long lifetime of the DTC order in the Floquet-Lindblad-Landau theory. These predictions have direct experimental relevance \cite{SM}. 

Our model can readily be generalized by taking into account the atomic motional degrees of freedom \cite{Domokos2002}, interactions between atoms \cite{Lewenstein2008}, local decoherence, and spontaneous emission \cite{DallaTorre2013,DallaTorre2016,Keeling2017}. In particular, our study raises an intriguing question of whether an intrinsically nonunitary DTC can possess absolute stability \cite{Keyserlingk2016} against arbitrary \emph{nonunitary} perturbation. Further studies along this line should give valuable hints for realizing a persistent DTC in the presence of realistic uncontrollable dissipation and decoherence. Another direction of research is to understand the Floquet-Lindblad spectra of other dynamical phases shown in Fig.~\ref{fig1}. We have already made some progress on the asymmetric DTC behavior \cite{SM}.

This work was supported by KAKENHI Grant No. JP26287088 from the Japan Society for the Promotion of Science, a Grant-in-Aid for Scientic Research on Innovative Areas ``Topological Materials Science" (KAKENHI Grant No. JP15H05855), and the Photon Frontier Network Program from MEXT of Japan, and the Mitsubishi Foundation. Z. G. was supported by MEXT. R. H. was supported by the Japan Society for the Promotion of Science through Program for Leading Graduate Schools (ALPS) and JSPS fellowship (JSPS KAKENHI Grant No. JP17J03189).

\bibliography{GZP_references}

\begin{thebibliography}{88}
\bibitem[{RS()}]{RS}
\bibinfo{note}{We have a finite resolution $0.001$, which is the length of the
  white stripes that separate adjacent phases. We do not rule out the
  possibility to find new phases in these narrow stripes.}
\end{thebibliography}

\clearpage
\widetext
\begin{center}
\textbf{\large Supplemental Materials}
\end{center}
\setcounter{equation}{0}
\setcounter{figure}{0}
\setcounter{table}{0}
\makeatletter
\renewcommand{\theequation}{S\arabic{equation}}
\renewcommand{\thefigure}{S\arabic{figure}}
\renewcommand{\bibnumfmt}[1]{[S#1]}

Here we provide the derivations of Eq.~(\ref{semicl}) in the main text, additional numerical results on the dynamical phases in the modulated open Dicke model, detailed derivation and analysis on the Floquet-Lindblad-Landau theory, and the details of the experimental implementations in cavity and circuit QED systems.

\section{Phase transition and semiclassical dynamics of the open Dicke model}

We briefly review some basic facts about the Dicke superradiant phase transition and derive the semiclassical equation of motion (Eq.~(\ref{semicl}) in the main text). 

\subsection{Quantum treatment for the Dicke phase transition}
We present the open-system counterpart of the method developed in Ref.~\cite{Brandes2003} which deals with the quantum phase transition in the isolated Dicke model. The same results have been obtained in Ref.~\cite{Carmichael2007} but basically using a semiclassical treatment. 

It is convenient to represent the system using the Holstein-Primakoff transformation:
\begin{equation}
\hat J_+=\hat b^\dag\sqrt{N-\hat b^\dag\hat b},\;\;\;\;
\hat J_-=\sqrt{N-\hat b^\dag\hat b}\hat b,\;\;\;\;
\hat J_z=\hat b^\dag\hat b-\frac{N}{2},
\label{HPT}
\end{equation} 
where the bosonic field operator $\hat b$ describes the atomic collective mode, which has a truncated Fock space up to $N$ bosons (corresponding to the fully spin-up polarized state $\ket{\Uparrow}\equiv\bigotimes^N_{j=1}\ket{\uparrow}$). Suppose that the system is in the symmetry-broken phase. We rewrite the photonic and atomic modes as $\hat a=\hat c+\alpha$ and $\hat b=\hat d-\beta$ ($|\alpha|,|\beta|\sim O(\sqrt{N})$) in the open Dicke model to obtain $\dot{\hat\rho}_t=-i[\hat H'(\lambda),\hat\rho_t]+\kappa\mathcal{D}[\hat c]\hat\rho_t$, where the Hamiltonian is given by
\begin{equation}
\begin{split}
\hat H'(\lambda)&=\omega\hat c^\dag\hat c+\omega_0\hat d^\dag\hat d+|\alpha|^2\omega+\left(|\beta|^2-\frac{N}{2}\right)\omega_0-\lambda\sqrt{1-\frac{|\beta|^2}{N}}(\alpha^*+\alpha)(\beta^*+\beta)\\
&+\left[\alpha\left(\omega-i\frac{\kappa}{2}\right)-(\beta^*+\beta)\sqrt{1-\frac{|\beta|^2}{N}}\lambda\right]\hat c^\dag
+\left\{\lambda(\alpha^*+\alpha)\sqrt{1-\frac{|\beta|^2}{N}}\left[1-\frac{(\beta^*+\beta)\beta}{2(N-|\beta|^2)}\right]-\beta\omega_0\right\}\hat d^\dag+{\rm H.c.}\\
&+\lambda(\alpha^*+\alpha)\sqrt{1-\frac{|\beta|^2}{N}}\left[\frac{\beta^*\hat d^2+\beta\hat d^{\dag2}}{2(N-|\beta|^2)}+\frac{(\beta^*+\beta)\hat d^\dag\hat d}{N-|\beta|^2}+\frac{(\beta^*+\beta)(\beta^*\hat d+\beta\hat d^\dag)^2}{8(N-|\beta|^2)^2}\right]\\
&+\lambda\sqrt{1-\frac{|\beta|^2}{N}}\left[1-\frac{(\beta^*+\beta)\beta^*}{2(N-|\beta|^2)}\right](\hat c^\dag+\hat c)\hat d+{\rm H.c.}.
\label{longHp}
\end{split}
\end{equation}
Here we have neglected the corrections of no more than $O(N^{-\frac{1}{2}})$ and used the gauge invariance of a general Lindblad equation $\dot{\hat\rho}_t=-i[\hat H,\hat\rho_t]+\sum_j\mathcal{D}[\hat L_j]\hat\rho_t$ under the transformations $\hat L_j\to\hat L_j+C_j$ and $\hat H\to\hat H+\sum_j\frac{i}{2}(C_j\hat L^\dag_j-C^*_j\hat L_j)$ \cite{Breuer2002}. To eliminate the linear terms with respect to the field operators in Eq.~(\ref{longHp}), we require the parameters $\alpha$ and $\beta$ to satisfy
\begin{equation}
\alpha\left(\omega-i\frac{\kappa}{2}\right)=(\beta^*+\beta)\sqrt{1-\frac{|\beta|^2}{N}}\lambda,\;\;\;\;
\lambda(\alpha^*+\alpha)\sqrt{1-\frac{|\beta|^2}{N}}\left[1-\frac{(\beta^*+\beta)\beta}{2(N-|\beta|^2)}\right]=\beta\omega_0,
\label{cond}
\end{equation}
which implies $\beta^*=\beta$ and
\begin{equation}
\beta\frac{\lambda^2}{\lambda^2_{\rm c}}\left(1-\frac{2\beta^2}{N}\right)=\beta,\;\;\;\;
\alpha=\frac{2\beta\lambda}{\omega-i\frac{\kappa}{2}}\sqrt{1-\frac{\beta^2}{N}},
\end{equation}
where the critical value $\lambda_{\rm c}$ reads
\begin{equation}
\lambda_{\rm c}=\frac{1}{2}\sqrt{\frac{\omega_0}{\omega}\left(\omega^2+\frac{\kappa^2}{4}\right)}.
\label{lamc}
\end{equation}
It is clear that there are nontrivial solutions ($|\alpha|,|\beta|\neq0$)
\begin{equation}
\beta^2=\frac{N}{2}(1-\mu),\;\;\;\;
|\alpha|^2=\frac{N\omega_0}{4\omega}(\mu^{-1}-\mu),\;\;\;\;
\mu\equiv\frac{\lambda^2_{\rm c}}{\lambda^2}
\label{nontrsol}
\end{equation}
if and only if $\lambda>\lambda_{\rm c}$ or equivalently $\mu<1$. Substituting Eqs.~(\ref{nontrsol}) and (\ref{cond}) into Eq.~(\ref{longHp}) yields
\begin{equation}
\hat H'=\omega\hat c^\dag\hat c+\frac{1+\mu}{2\mu}\omega_0\hat d^\dag\hat d+\frac{(3+\mu)(1-\mu)}{8\mu(1+\mu)}\omega_0(\hat d+\hat d^\dag)^2+\lambda\mu\sqrt{\frac{2}{1+\mu}}(\hat c^\dag+\hat c)(\hat d^\dag+\hat d)-\frac{N(1+\mu^2)+1-\mu}{4\mu}\omega_0.
\label{simpHp}
\end{equation}

It is known \cite{Brandes2003} that without photon loss ($\kappa=0$), the ground state of $\hat H'$ (\ref{simpHp}) is a squeezed (including both single-mode and two-mode squeezed) vacuum with respect to $\hat c$ and $\hat d$. In the presence of photon loss, the steady state becomes not only squeezed but also mixed. However, since $\hat c$ and $\hat d$ are obtained by a large translation of the order of $O(\sqrt{N})$ from $\hat a$ and $\hat b$, the expectation values of local observables (e.g., single-atom spin polarization) in the thermodynamic limit are expected to coincide with those of $\ket{\alpha}\otimes\ket{-\beta}$, i.e., the direct product of photon and atomic-spin coherent states.

\subsection{Heisenberg equation of motion and its semiclassical reduction}
To work out the semiclassical equation of motion, we first consider the Heisenberg equation of motion. For a general \emph{time-dependent} Lindblad equation $\dot{\hat\rho}_t=\mathcal{L}_t\hat\rho_t=-i[\hat H(t),\hat\rho_t]+\sum_j\mathcal{D}[\hat L_j(t)]\hat\rho_t$, the open-system Heisenberg equation for an observable $\hat O$ (explicitly time-independent) is given by \cite{Breuer2002}
\begin{equation}
\frac{d\langle\hat O\rangle}{dt}=\langle\mathcal{L}^\dag_t\hat O\rangle=\left\langle i[\hat H(t),\hat O]+\sum_j\left(\hat L^\dag_j(t)\hat O\hat L_j(t)-\frac{1}{2}\{\hat L^\dag_j(t)\hat L_j(t),\hat O\}\right)\right\rangle,
\label{LinHei}
\end{equation}
where $\langle...\rangle\equiv{\rm Tr}[...\hat\rho_t]$ is the instantaneous ensemble average. Applying Eq.~(\ref{LinHei}) to the modulated open Dicke model and using the commutation relations $[\hat a,\hat a^\dag]=1$ and $[\hat J_\mu,\hat J_\nu]=i\epsilon_{\mu\nu\sigma}\hat J_\sigma$, we obtain
\begin{equation}
\begin{split}
\frac{d\langle\hat a\rangle}{dt}&=-\left(i\omega+\frac{\kappa}{2}\right)\langle\hat a\rangle-\frac{2i\lambda_t}{\sqrt{N}}\langle\hat J_x\rangle\\
\frac{d\langle\hat J_x\rangle}{dt}=-\omega_0\langle\hat J_y\rangle,\;\;\;\;
\frac{d\langle\hat J_y\rangle}{dt}&=\omega_0\langle\hat J_x\rangle-\frac{2\lambda_t}{\sqrt{N}}\langle(\hat a+\hat a^\dag)\hat J_z\rangle,\;\;\;\;
\frac{d\langle\hat J_z\rangle}{dt}=\frac{2\lambda_t}{\sqrt{N}}\langle(\hat a+\hat a^\dag)\hat J_y\rangle.
\end{split}
\label{DFLH}
\end{equation}
We can also show $\frac{d\langle\hat J^2\rangle}{dt}=0$ ($\hat J^2\equiv\sum_\mu\hat J^2_\mu$) from $[\hat J^2,\hat J_\mu]=0$, i.e., $\langle\hat J^2\rangle$ is conserved. In particular, if the atoms are initialized (by, e.g., optical pumping) to be a fully spin-polarized state like $\ket{\Downarrow}\equiv\bigotimes^N_{j=1}\ket{\downarrow}$, which is the case in real experiments \cite{Esslinger2010,Parkins2014}, we have $\langle\hat J^2\rangle=\frac{N}{2}(\frac{N}{2}+1)$.

At the mean-field level, we approximate $\langle\hat a\hat J_\mu\rangle$ as $\langle\hat a\rangle\langle\hat J_\mu\rangle$, which has an order of magnitude $O(N^{\frac{3}{2}})$ while the quantum fluctuation is expected to be no more than $O(N)$. Such an approximation should become exact in the thermodynamic limit. In terms of the scaled variables $\tilde\alpha\equiv\frac{\langle\hat a\rangle}{\sqrt{N}}$ and $j_\mu\equiv\frac{\langle\hat J_\mu\rangle}{N}$, Eq.~(\ref{DFLH}) gives a closed set of semiclassical equations:
\begin{equation}
\frac{d\tilde\alpha}{dt}=-\left(i\omega+\frac{\kappa}{2}\right)\tilde\alpha-2i\lambda_tj_x,\;\;\;\;
\frac{dj_x}{dt}=-\omega_0j_y,\;\;\;\;
\frac{dj_y}{dt}=\omega_0j_x-2\lambda_t(\tilde\alpha+\tilde\alpha^*)j_z,\;\;\;\;
\frac{dj_z}{dt}=2\lambda_t(\tilde\alpha+\tilde\alpha^*)j_y,
\label{alphasemicl}
\end{equation}
which can finally be rewritten in the form of Eq.~(\ref{semicl}) in the main text after the substitution $\tilde\alpha=\sqrt{\frac{\omega}{2}}x+\frac{ip}{\sqrt{2\omega}}$. Note that $j^2\equiv\sum_\mu j^2_\mu$ continues to be a conserved quantity in Eq.~(\ref{alphasemicl}), which takes on the value of $\frac{1}{4}$ when $N\to\infty$. This is why the trajectory of atomic angular momenta is confined on the Bloch sphere. It is also worth mentioning that for a large but finite $N$ we can systematically calculate the corrections of the order of $N^{-k}$ by means of the cumulant expansion \cite{Keeling2017}, which is beyond the scope of the present Letter and we would like to leave it for future work.

Finally, we emphasize that once the semiclassical dynamics becomes chaotic, it is, in practice, impossible to obtain exact numerical results after a short time interval. This is due to the \emph{exponential} amplification of inevitable numerical errors. Nevertheless, we can still observe qualitative behaviors of irregular trajectories covering areas, based on which we judge that the system is in the thermal phase.

\section{Diagnosis and interpretation of the dynamical phases in the modulated open Dicke model}
In this section we provide a useful tool to diagnose all kinds of dynamical phases presented in Fig.~\ref{fig2} in the main text. This approach is inspired by the recent studies on \emph{synchronization} \cite{Tony2013,Lesanovsky2015,Nunnenkamp2017} in nonlinear classical and quantum systems. With the help of this tool, we identify several novel dynamical phases beyond those discussed in the main text. The rich dynamical phase diagram can systematically be understood from bifurcation theory.

\subsection{Synchronization-based approach to diagnosing dynamical phases}
\label{Synch}
The rigidity of the DTC order in isolated system is usually explained as a result of many-body synchronization. As for the Dicke model, while we do have long-range (actually all to all) interactions between atoms mediated by photons \cite{Esslinger2010}, it is more convenient to regard the atoms as a rotor that couples nonlinearly to a harmonic oscillator that represents single-mode photons. In this picture, we can quantify the degree of synchronization through the phase difference between the rotor and the oscillator. 

Denoting the phase of the photons as $\phi_a$ and that of the atoms as $\phi_b$, we have\begin{equation}
\langle\hat a\rangle\equiv|\langle\hat a\rangle|e^{i\phi_a},\;\;\;\;
\langle\hat b\rangle\equiv|\langle\hat b\rangle|e^{i\phi_b},
\end{equation}
where $\hat b$ is the bosonic mode of collective atomic excitations defined in Eq.~(\ref{HPT}). In the thermodynamic limit, the phases can semiclassically be evaluated through the relations
\begin{equation}
\phi_a={\rm Arg}\left(\sqrt{\omega}x+\frac{ip}{\sqrt{\omega}}\right),\;\;\;\;
\phi_b={\rm Arg}(j_x+ij_y).
\end{equation}
Note that $\phi_a-\phi_b=\pi\mod2\pi$ for the symmetry-broken ground states of the isolated Dicke model, and that it slightly deviates from $\pi$ for a small nonzero $\kappa$. 

\begin{figure}
\begin{center}
       \includegraphics[width=16cm, clip]{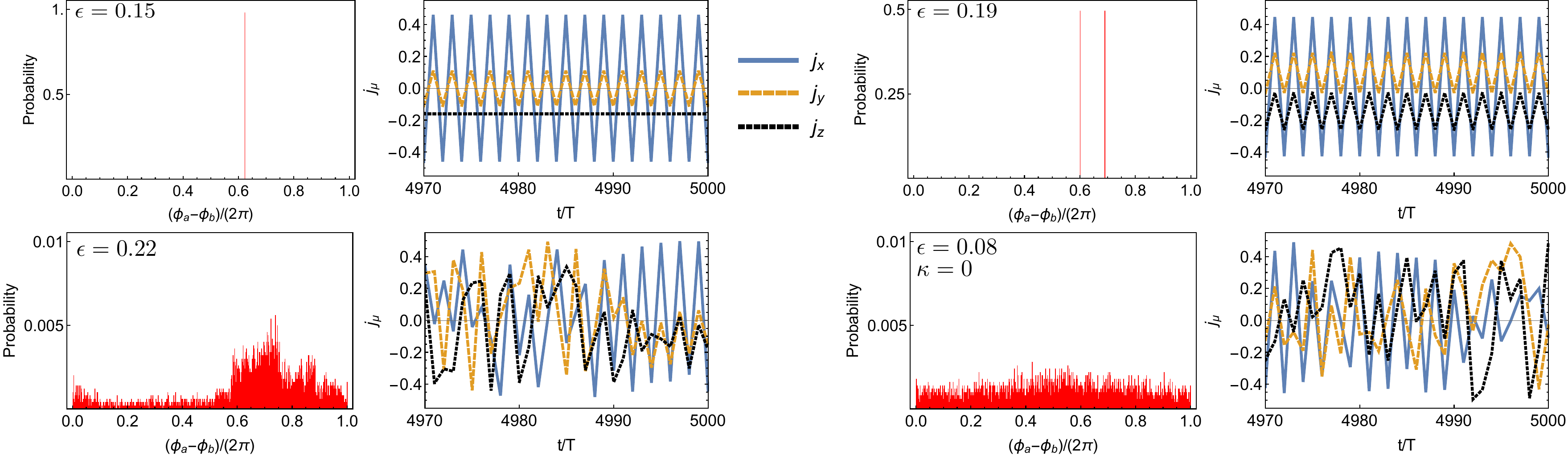}
       \end{center}
   \caption{Probability distribution of the phase difference (PDPD) $\phi_a-\phi_b$ over $5000$ periods and the stroboscopic dynamics of atomic angular momenta in the last $30$ periods for the normal DTC phase (top left), the asymmetric period-doubling phase (top right) and the thermal phases (bottom). We use $\kappa=0.25$ unless indicated otherwise. A single peak in the PDPD splits into two when the doublet becomes asymmetric. For thermal phases, the PDPD spreads almost uniformly in the isolated limit ($\kappa=0$) while inhomogeneously for a nonzero $\kappa$.}
   \label{figS1}
\end{figure}

Let us consider the modulated open Dicke model in the normal DTC phase (symmetric doublet). The states of the system at the end of an odd number of periods and that of an even number of periods are exactly related to each other by the parity operator, implying $\phi_a\to\phi_a+\pi\mod2\pi$ and $\phi_b\to\phi_b+\pi\mod2\pi$ after each period. Therefore, the photon and atomic phases are perfectly synchronized in the normal DTC phase, as indicated by a single peak in the probability distribution of the phase difference (PDPD) $\phi_a-\phi_b$ calculated at $t_n=nT$ ($n=1,2,...,5000$). We give an example in the top left two panels in Fig.~\ref{figS1}.

When the period doubling becomes asymmetric, the PDPD splits into two peaks (see the top right two panels in Fig.~\ref{figS1}), since the increments of $\phi_a$ and $\phi_b$ after each period are no long the same, although those after every \emph{two} periods are both $2\pi$. When the system enters the thermal phase, the PDPD spreads to everywhere over $[0,2\pi)$, implying the loss of synchronization (see the panels in the bottom in Fig.~\ref{figS2}). The PDPD also becomes continuous for a symmetric limit-cycle pair, but is localized in a finite range with singularities at the boundaries (see the second column from the left in Fig.~\ref{figS2}). We can see that the behavior of the PDPD sharply distinguishes different dynamical phases.

\subsection{Understanding the dynamical phases from bifurcation theory}
The richness of the dynamical phases in the modulated open Dicke model arises from the nonlinearity of the semiclassical dynamics (\ref{alphasemicl}) which, in turn, originates from the finite-level nature of the atomic spectrum. Here by the finite-level nature, we mean that the atomic excitations can be saturated for a given $N$. Such saturation effects become increasingly more significant and hence survive even in the thermodynamic limit as the system is excited farther away from the steady state. This is precisely the case with the modulated open Dicke model.

\begin{figure}
\begin{center}
        \includegraphics[width=8cm, clip]{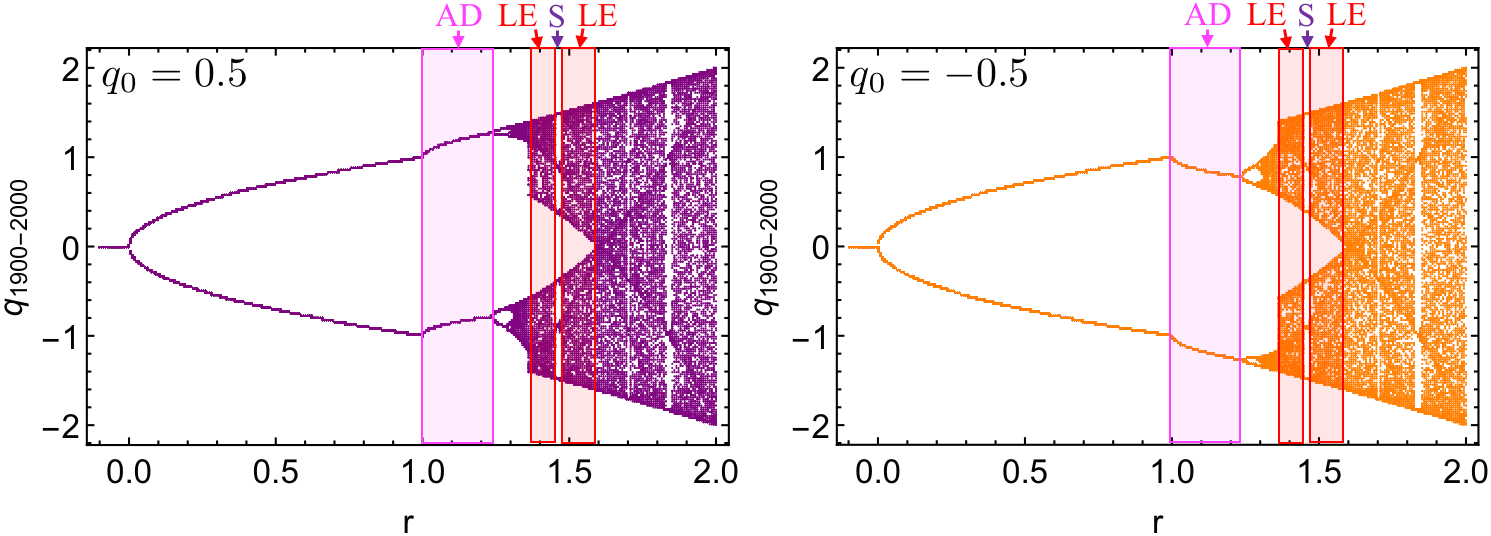}
      \end{center}
   \caption{(color online). Steady orbit $q_{1900-2000}$ of the recurrence equation (\ref{iter}) starting from $q_0=0.5$ (left panel) and $-0.5$ (right panel) for different parameter $r$. Besides the symmetric doublet for $0<r<1$, we observe the asymmetric doublet (indicated as AD) and even the period sextupling (S) embedded in the locally ergodic phase (LE).}
   \label{fig3S}
\end{figure}

In the language of nonlinear dynamical systems \cite{Holmes1983}, different dynamical phases are caused by certain kinds of bifurcations. The normal DTC order is essentially an interplay of pitchfork bifurcation and parity symmetry. It appears already in a simple recurrence series:
\begin{equation}
q_{n+1}=-(r+1)q_n+q^3_n,
\label{iter}
\end{equation}
which is a combination of a minimal discrete dynamics $q\to(r+1)q-q^3$ for supercritical pitchfork bifurcation and the inversion $q\to-q$. Note that this is neither a supercritical pitchfork bifurcation alone, after which the dynamics converges to one of the fixed points, nor a subcritical pitchfork bifurcation, after which no stable fixed point exists. As shown in Fig.~\ref{fig3S}, we see symmetric period doubling for $0<r<1$. When $r$ exceeds $1$, the symmetric orbit becomes unstable and a local period-doubling bifurcation, which is well-known in the logistic map \cite{May1976}, occurs at the ensemble level, leading to the asymmetric period doubling at the trajectory level. For larger $r$, we observe a narrow sextet window embedded in the locally (two-branched) ergodic phase, resembling the period-sextupling phase sandwiched by limit-cycle pairs shown in the top row of Fig.~\ref{fig2} in the main text. While the toy model (\ref{iter}) and the modulated Dicke model share many features, only the latter shows transitions between limit cycles and fixed points around $\epsilon=0$ and $-0.07$. These transitions are a Floquet version of the Hopf bifurcation \cite{Xiao2007}, which requires at least two continuous variables.

\begin{figure}
\begin{center}
       \includegraphics[width=18cm, clip]{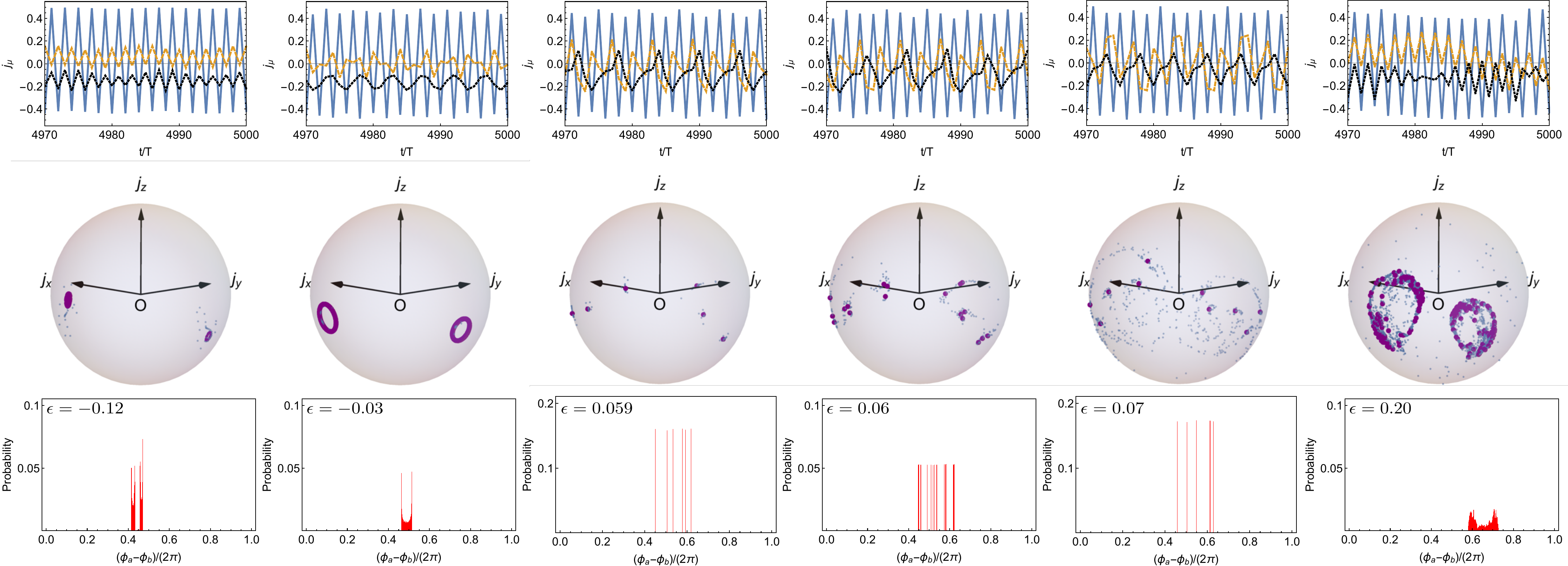}
       \end{center}
   \caption{Stroboscopic dynamics of the atomic angular momenta over the last $30$ periods (top), the phase-space trajectories of the last $200$ periods projected onto the pseudospin Bloch sphere (middle) and the corresponding probability distribution of the phase difference (bottom). The other parameters are chosen to be $\kappa=0.25$ and $\lambda=1$. In addition to a limit-cycle pair ($\epsilon=-0.03$), we find new dynamical phases including an asymmetric limit-cycle pair ($\epsilon=-0.12$), an asymmetric sextet ($\epsilon=0.059$), ten- ($\epsilon=0.07$) and even eighteen-fold ($\epsilon=0.06$) multiplets, and a locally ergodic phase ($\epsilon=0.2$).}
   \label{figS2}
\end{figure}

\subsection{Higher multiplets, asymmetric limit-cycle pair and locally ergodic phase}
We apply the powerful approach introduced in Sec.~\ref{Synch} to explore the dynamical phases under the parameter choices $\kappa=0.25$ and $\lambda=1$. Typical numerical results are presented in Fig.~\ref{figS2}. For $\epsilon=-0.12$, we find a pair of asymmetric limit cycles, indicated by two continuous compact regions in the PDPD. We expect that this phase results from a Hopf bifurcation of an asymmetric doublet order. For $\epsilon=0.06$ ($0.07$), we find that the period of the stroboscopic dynamics becomes eighteen-fold (ten-fold) in an asymmetric (symmetric) manner, which can be read out from the number of peaks in the PDPD. While the ten-fold dynamics should be a basic cycle \cite{May1976} like the sextet order, the eighteen-fold dynamics might be a novel bifurcation from an asymmetric sextet order observed for $\epsilon=0.059$. Furthermore, when $\epsilon=0.2$, we observe a locally ergodic phase where the trajectories cover two separated 
 \emph{areas} on the angular-momenta sphere (in the quadrature plane). This phase is more chaotic than a limit-cycle pair, where the trajectories are one dimensional, yet less chaotic than a thermal phase, where there is only a single 
 area covered by the trajectories and the DTC order is destroyed. These features are well captured by the PDPD, which is still localized but the boundary singularities are smeared out.

\section{Further numerical results for the few-qubit Dicke model}
In this section, we show how the transient DTC behavior can be understood from the spectra of the Floquet-Lindblad superoperators and present additional numerical results for the three-qubit modulated open Dicke model.

\subsection{Understanding the DTC behavior from the Floquet-Lindblad spectrum}
As shown in the main text, the transient DTC behavior of the modulated open Dicke model in the deep quantum regime emerges only in the presence of both strong coupling and dissipation. Here we present further discussions on this issues based on the spectrum analysis of the Floquet-Lindblad superoperator $\mathcal{U}_{\rm F}=\mathcal{T}e^{\int_0^Tdt\mathcal{L}(\lambda_t)}$.

Let $\{u_\alpha\}_{\alpha}$ and $\{|u_\alpha)\}_\alpha$ be a set of eigenvalues and that of right (super)eigenvectors of $\mathcal{U}_{\rm F}$, respectively.
The vector $|\rho_0)$ representing the initial state $\hat{\rho}_0$ evolves stroboscopically as
\begin{equation} 
|\rho_{nT})=\sum_\alpha u_\alpha^n(\tilde u_\alpha|\rho_0)|u_\alpha), 
\end{equation}
where $|\tilde u_\alpha)$ is the left eigenvector corresponding to $|u_\alpha)$ and the Hilbert-Schmidt inner product is defined as $(A|B)\equiv{\rm Tr}[\hat A^\dag\hat B]$. The eigenvector $|u_+)$ with $u_+=1$ represents the stationary state of the Floquet-Lindblad dynamics.
If we assume that the stationary state is unique, the other eigenvectors have eigenvalues with $|u_\alpha|<1$ and decay exponentially due to the factor $u_\alpha^n$.
However, if there exists a single eigenvector $|u_-)$ whose eigenvalue $u_-$ is close to $-1$, 
$|\rho_{nT})$ becomes a mixture of two eigenmodes of $|u_\pm)$ for relatively large $n$.
In this case, the state can be approximated as $|\rho_{nT})\simeq c_+|u_+)+c_-|u_-)$, leading to \begin{equation}
|\rho_{(n+1)T})\simeq c_+|u_+)-c_-|u_-),\;\;\;\;|\rho_{(n+2)T})\simeq c_+|u_+)+c_-|u_-),
\label{upm}
 \end{equation}
if we neglect the decay of $|u_-)$. This regime exhibits the DTC order with a period of $2T$. Note that $c_+=1$ if $|u_+)$ is the \emph{normalized} steady state with the unit trace. 

\begin{figure}
\begin{center}
\includegraphics[width=10.98cm]{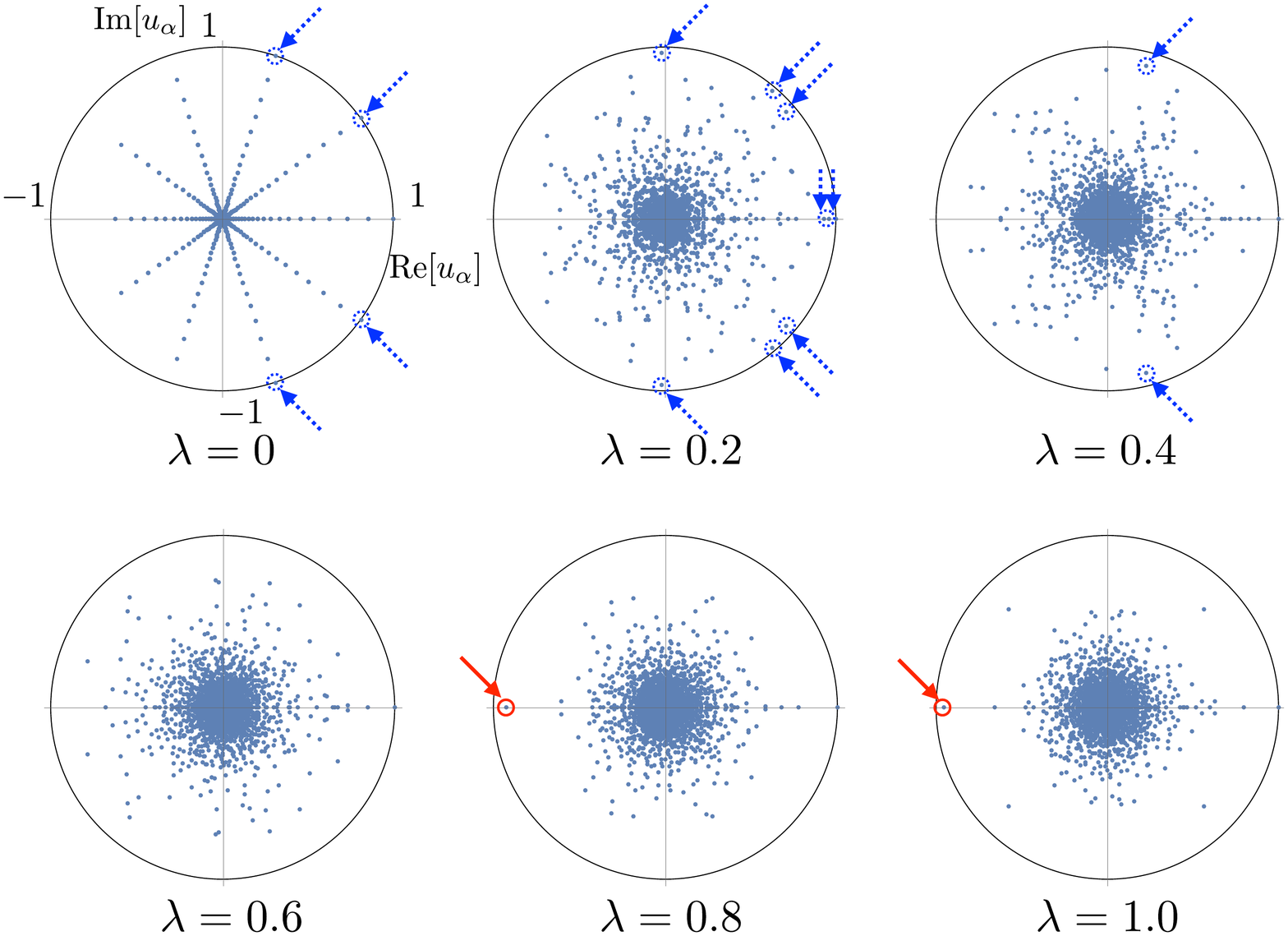}
\caption{Spectra of the Floquet-Lindblad superoperator $\mathcal{U}_{\rm F}$ for different coupling strength $\lambda=0, 0.2, 0.4, 0.6$ and $0.8$.
Arrows show long-lived eigenmodes whose eigenvalues satisfy $|u_\alpha|\geq 0.9$ and $u_\alpha\neq 1$.
Note that such an eigenmode does not exist for $\lambda=0.6$.
As indicated by the dotted arrows, such eigenmodes are not close to $-1$ in the weak-coupling regime ($\lambda\leq 0.6$).
On the other hand, in the strong-coupling regime ($\lambda\geq 0.8$), there exists a single real eigenmode whose eigenvalue is close to $-1$, as indicated by the solid arrows.
The data are obtained by the exact diagonalization method, where we truncate the Hilbert space up to $16$ photons.}
\label{figS3}
\end{center}
\end{figure}

It is worthwhile to mention that the DTC order manifests only if we look at an odd-parity observable $\hat O$, such as $\hat x$, $\hat p$, $\hat J_x$ and $\hat J_y$. That is, $\mathcal{P}\hat O\equiv\hat P\hat O\hat P=-\hat O$, where $\mathcal{P}$ is the parity superoprator. Note that $(1|\mathcal{P}|u)={\rm Tr}[\hat P^2\hat u]=(1|u)$, so the unique steady state $|u_+)$ must feature even parity, i.e., $\mathcal{P}|u_+)=|u_+)$, and each odd-parity operator must be traceless. We can argue that $|u_-)$ is an odd-parity operator from the perspective of continuous deformation of $\mathcal{U}_{\rm F}$ from the ideal form $\mathcal{P}e^{\mathcal{L}(\lambda)\frac{T}{2}}$ and the discrete nature of parity eigenvalues. Therefore, $|u_+)$ contributes nothing to $\langle\hat O\rangle$ and only $|u_-)$ in Eq.~(\ref{upm}) contributes a finite expectation value and thus gives rise to the DTC order. On the other hand, the expectation of an even-parity operator like $\hat n=\hat a^\dag\hat a$ and $\hat J_z$ stays unchanged during the stroboscopic evolution given by Eq.~(\ref{upm}), since only $|u_+)$ contributes a finite value but the coefficient does not flip its sign after a period.

Figure \ref{figS3} shows the spectra of $\mathcal{U}_F$ for different coupling strength $\lambda$. The other parameters are fixed to be $\kappa=0.05$ and $\epsilon=0.1$, which are the same as those in the main text. Arrows in Fig.~\ref{figS3} show long-lived eigenmodes whose eigenvalues satisfy $|u_\alpha|\geq 0.9$ and $u_\alpha\neq 1$. As indicated by the dotted arrows, such eigenmodes are not close to $-1$ in the weak-coupling regime ($\lambda\leq 0.6$). We note that the approximated ten-fold rotation symmetry of the spectrum for $\lambda=0$ is due to the specific choice $\epsilon=0.1$ ($\epsilon^{-1}=10$). On the other hand, in the strong-coupling regime ($\lambda\geq 0.8$), there exists a single real eigenmode whose eigenvalue is close to $-1$, as indicated by solid arrows. This eigenmode corresponds to $|u_-)$ above and contributes to the transient DTC order.

\subsection{Numerical results for $N=3$}
In the main text, we have seen that, in the strong-coupling regime, the transient DTC order is stabilized by dissipation even for two qubits. We here demonstrate that the same feature is shared by the modulated open Dicke models with $N=3$.

\begin{figure}
\begin{center}
\includegraphics[width=8cm]{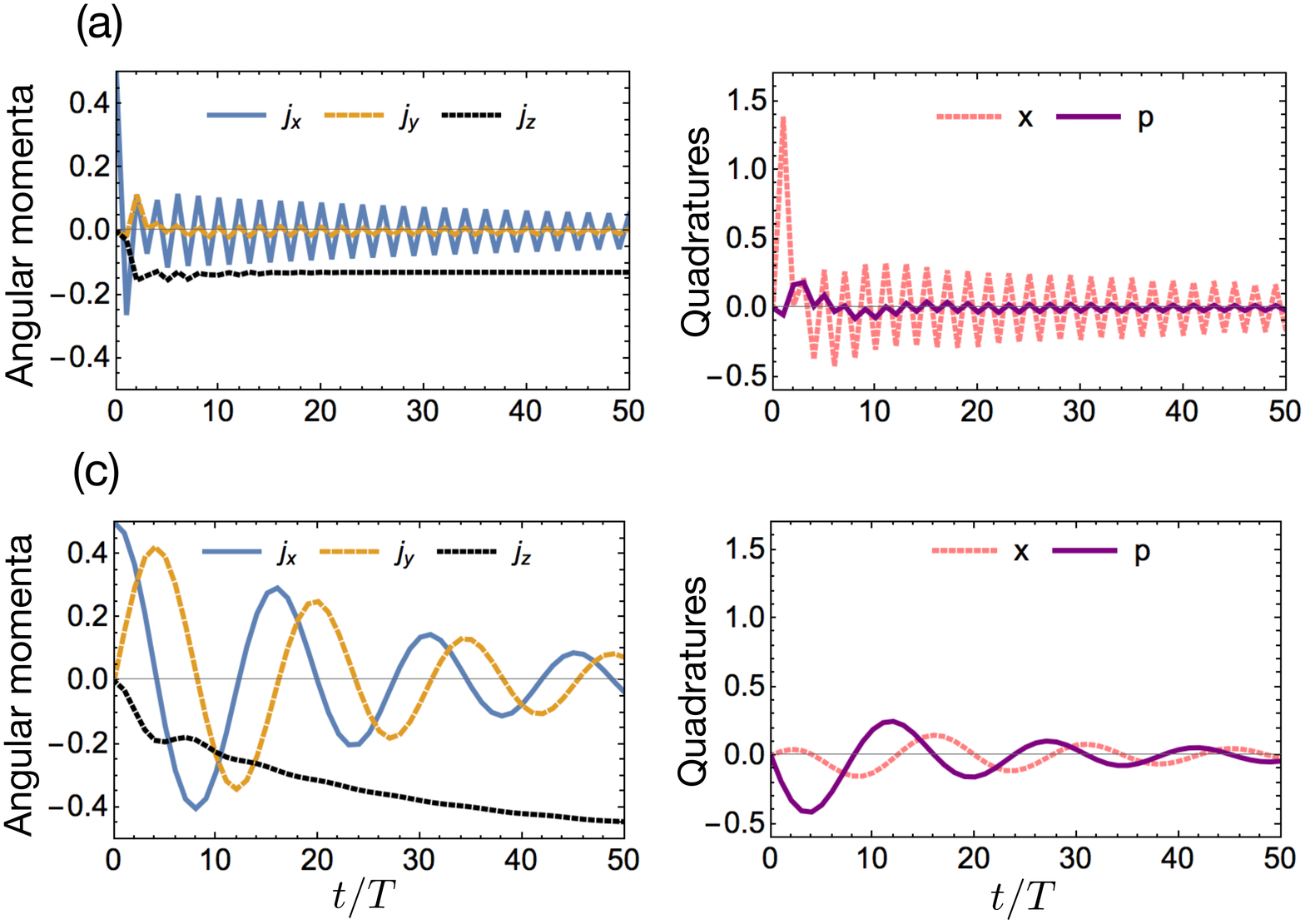}
\includegraphics[width=8cm]{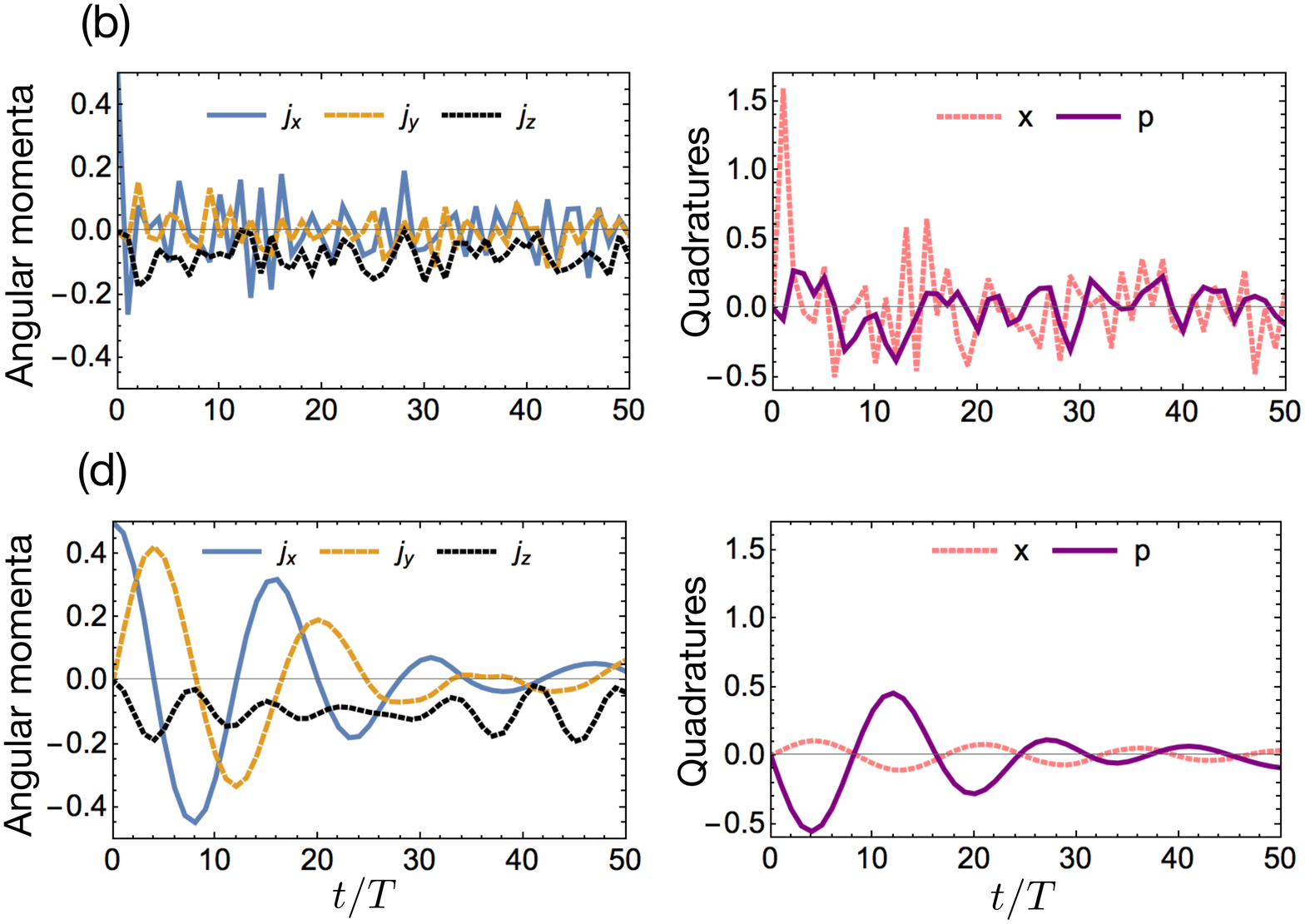}
\caption{(color online).
(a) Stroboscopic dissipative dynamics of the three-qubit Dicke model with $\kappa=0.05, \epsilon=0.05$ and $\lambda=1$ (strong coupling).
Scaled angular momenta of $j_x$ (solid), $j_y$ (dashed), $j_z$ (dotted), quadratures $x$ (dotted) and $p$ (solid) are shown.
We clearly see that the DTC order appears.
(b) Stroboscopic dynamics for isolated systems ($\kappa=0, \epsilon=0.05$, and $\lambda=1$) in the strong coupling regime.
(c) Stroboscopic dissipative dynamics in the weak-coupling regime ($\kappa=0.05, \epsilon=0.05$ and $\lambda=0.1$).
(d) Stroboscopic unitary dynamics in the weak-coupling regime ($\kappa=0, \epsilon=0.05$ and $\lambda=0.1$).
In (b)-(d), no DTC order appears.
The initial state is always chosen to be $\ket{\Rightarrow}\otimes \ket{0}$, where $\ket{\Rightarrow}\equiv\bigotimes^N_{j=1}\ket{\rightarrow}$ is the eigenstate of $\hat{J}_x$ with eigenvalue $N/2\:(N=3)$ and $\ket{0}$ is the photon vacuum.
}
\label{figS4}
\end{center}
\end{figure}

Figure \ref{figS4} shows time evolutions of the scaled angular momenta and quadratures for three-qubit Dicke models with different parameters (the detuning is fixed to be $\epsilon=0.05$).
As shown in Fig. \ref{figS4} (a), the transient DTC order emerges for the dissipative case $(\kappa=0.05)$ in the strong-coupling regime ($\lambda=1$).
On the other hand, no transient DTC order appears in the isolated counterpart ($\kappa=0$ and $\lambda=1$), as shown in Fig. \ref{figS4} (b).
In the weak-coupling regime ($\lambda=0.1$), the dissipative ($\kappa=0.05$, Fig. \ref{figS4} (c)) and isolated ($\kappa=0$, Fig. \ref{figS4} (c)) cases behave similarly to each other, both of which do not exhibit the DTC order.

\section{Details of the Floquet-Lindblad-Landau theory}
In this section we give a detailed derivation and analysis of the Floquet-Lindblad-Landau theory. A possible phenomenology for the asymmetric DTC behavior is also discussed.

\subsection{Derivation from the open Dicke model}
We first derive the effective dynamics of the photon degree of freedom on the basis of 
a semiclassical argument. The semiclassical equation of motion is given by
\begin{equation}
\dot p=-\omega^2 x-\frac{\kappa}{2}p-2\lambda\sqrt{2\omega}j_x,\;\;\;
\dot x=p-\frac{\kappa}{2}x,\;\;\;\;\;
\dot j_x=-\omega_0j_y,\;\;\;
\dot j_y=\omega_0j_x-2\lambda\sqrt{2\omega}xj_z,\;\;\;
\dot j_z=2\lambda\sqrt{2\omega}xj_y.
\label{semicl2}
\end{equation}
Assuming that $\omega_0\gg\omega$, we expect that the atomic degrees of freedom will soon equilibrate ($\dot j_\mu=0$, $\mu=x,y,z$) upon a small change in the photon degree of freedom. In this case, $j_x$ can be estimated from $x$ via
\begin{equation}
j_x=-\frac{\lambda\sqrt{2\omega} x}{\sqrt{\omega^2_0+8\omega\lambda^2x^2}}.
\label{stdjx}
\end{equation}
Note that the minus sign comes from the assumption that $j_z<0$, which can be justified by a low-energy atomic state like $j_z=-\frac{1}{2}$ at the initial time. Substituting Eq.~(\ref{stdjx}) into the left two equations of Eq.~(\ref{semicl2}) yields a closed equation of motion in terms of $x$ and $p$ alone: 
\begin{equation}
\dot p=-\omega^2 x-\frac{\kappa}{2}p+\frac{4\lambda^2\omega x}{\sqrt{\omega^2_0+8\omega\lambda^2x^2}},\;\;\;\;
\dot x=p-\frac{\kappa}{2}x.
\label{eomxp}
\end{equation}
If $8\omega\lambda^2x^2\ll\omega^2_0$, which turns out to be equivalent to $\lambda\simeq\lambda_{\rm c}$ (since $x\sim\frac{\lambda\omega_0}{\lambda^2_{\rm c}\sqrt{\omega}}\sqrt{1-\frac{\lambda^4_{\rm c}}{\lambda^4}}$), Eq.~(\ref{eomxp}) can well be approximated by
\begin{equation}
\dot p=-\omega^2 x-\frac{\kappa}{2}p+\frac{4\lambda^2\omega}{\omega_0}x-\frac{16\lambda^4\omega^2}{\omega^3_0}x^3,\;\;\;\;
\dot x=p-\frac{\kappa}{2}x,
\end{equation}
from which we can infer that the Lindblad master equation should be given by
\begin{equation}
\dot{\hat\rho}_t=-i\left[\omega\hat a^\dag\hat a-\frac{\lambda^2}{\omega_0}(\hat a^\dag+\hat a)^2+\frac{\lambda^4}{\omega^3_0N}(\hat a^\dag+\hat a)^4,\hat\rho\right]+\kappa\mathcal{D}[\hat a]\hat\rho_t.
\label{DFLG}
\end{equation}
By replacing $\lambda$ with $\lambda_{t}=\lambda_{t+T}$, the above equation describes the dissipative and Floquet counterpart of the well-known Landau theory. In this sense, while derived from the open Dicke model, the general form of Eq.~(\ref{DFLG}) 
\begin{equation}
\dot{\hat\rho}_t=-i[\hat H_{\rm L},\hat\rho_t]+\kappa\mathcal{D}[\hat a]\hat\rho_t,\;\;\;\;
\hat H_{\rm L}=\omega\hat a^\dag\hat a-\frac{\Omega_2}{4}(\hat a^\dag+\hat a)^2+\frac{\Omega_4}{32N}(\hat a^\dag+\hat a)^4
\label{GFLL}
\end{equation}
should widely be applicable to periodically driven single-mode open quantum systems. It is worth mentioning that Eq.~(\ref{GFLL}) features a parity symmetry with respect to $\hat P_a=e^{i\pi\hat a^\dag\hat a}$. 

\begin{figure}
\begin{center}
        \includegraphics[width=10cm, clip]{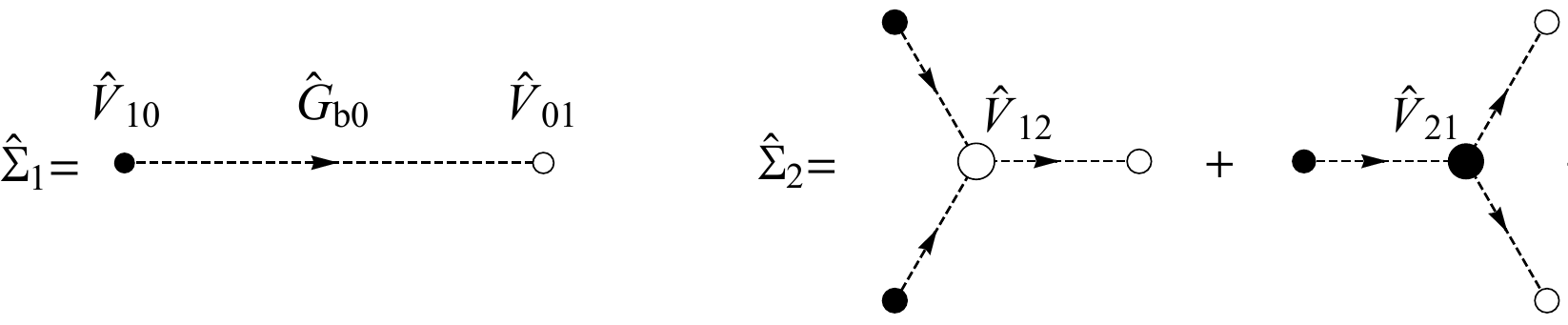}
      \end{center}
   \caption{Diagramatic illustration of the single (left) and double (right) atomic excitation virtual processes. The creation, annihilation and free propagation of an atomic excitation is represented are a filled dot, an open dot and a dashed line, respectively. The larger open (filled) dot refers to the coalescence of two atomic excitations (the split of an atomic excitation).}
      \label{figFD}
\end{figure}

A result consistent with Eq.~(\ref{DFLG}) can be obtained by using adiabatic elimination, which is a purely quantum treatment. To do this, we first write down the Dicke Hamiltonian
\begin{equation}
\hat H_{\rm Dicke}=\omega\hat a^\dag\hat a+\omega_0\hat b^\dag\hat b+\lambda(\hat a^\dag+\hat a)\left(\hat b^\dag\sqrt{1-\frac{\hat b^\dag\hat b}{N}}+\sqrt{1-\frac{\hat b^\dag\hat b}{N}}\hat b\right)-\frac{N}{2}\omega_0,
\end{equation}
which can be approximated as
\begin{equation}
\hat H=\omega\hat a^\dag\hat a+\omega_0\hat b^\dag\hat b+\lambda(\hat a^\dag+\hat a)(\hat b^\dag+\hat b)-\frac{\lambda}{2N}(\hat a^\dag+\hat a)(\hat b^{\dag2}\hat b+\hat b^\dag\hat b^2)
\end{equation}
if $\hat b^\dag\hat b\ll N$, i.e., almost all the atoms are at their ground states. In the case of $\omega\ll\omega_0$ and starting from the ground state $|0\rangle_b$ of the atomic ensemble, the creation of an atomic excitation is expected to be blocked by a large energy discrepancy. The task of adiabatic elimination is nothing but to find out an effective Hamiltonian $\hat H_a$ of the photon field alone, which satisfies
\begin{equation}
e^{-i\hat H_a t}\simeq{}_b\langle0|e^{-i\hat Ht}|0\rangle_b.
\end{equation}
This problem is more conveniently solved in the frequency domain, i.e., by using the Green's-function formalism and the Dyson equation 
\begin{equation}
\hat G^{-1}(\Omega)=\hat G^{-1}_0(\Omega)-\hat\Sigma(\Omega),
\end{equation}
where $\hat G(\Omega)={}_b\langle0|(\Omega-\hat H+i0^+)^{-1}|0\rangle_b$ is the photon Green's function in the presence of an interaction, $\hat G_0(\Omega)=(\Omega-\omega\hat a^\dag\hat a+i0^+)^{-1}$ is the free photon Green's function and $\hat\Sigma(\Omega)$ is the self-energy, which can perturbatively be computed by summing up the contribution from irreducible virtual processes (see Fig.~\ref{figFD}). The leading-order contribution arises from the single atomic excitation virtual process and is given by
\begin{equation}
\hat\Sigma_1(\Omega)={}_b\langle0|\hat V_{01}\hat G_{b0}(\Omega)\hat V_{10}|0\rangle_b=\frac{\lambda^2}{\Omega-\omega_0}(\hat a^\dag+\hat a)^2,
\label{Sigma1}
\end{equation}
where $\hat V_{10}=\hat V^\dag_{01}=\lambda(\hat a^\dag+\hat a)\hat b^\dag$ and $\hat G_{b0}(\Omega)=(\Omega-\omega_0\hat b^\dag\hat b+i0^+)^{-1}$ is the free Green's function of atoms. Note that in Eq.~(\ref{Sigma1}) $i0^+$ is neglected since typically $\Omega\ll\omega_0$. The subleading-order contribution arises from the two atomic excitation virtual process and is given by
\begin{equation}
\begin{split}
\hat\Sigma_2(\Omega)&={}_b\langle0|(\hat V_{01}\hat G_{b0}(\Omega)\hat V_{12}\hat G_{b0}(\Omega)\hat V_{10}\hat G_{b0}(\Omega)\hat V_{10}+\hat V_{01}\hat G_{b0}(\Omega)\hat V_{01}\hat G_{b0}(\Omega)\hat V_{21}\hat G_{b0}(\Omega)\hat V_{10})|0\rangle_b\\
&=-\frac{2\lambda^3}{N(\Omega-\omega_0)^2(\Omega-2\omega_0)}(\hat a^\dag+\hat a)^4,
\end{split}
\end{equation}
where $\hat V_{21}=\hat V^\dag_{12}=-\frac{\lambda}{2N}(\hat a^\dag+\hat a)\hat b^{\dag2}\hat b$ and the factor of $2$ in the numerator results from ${}_b\langle0|\hat b^2\hat b^{\dag2}\hat b\hat b^\dag|0\rangle_b={}_b\langle0|\hat b\hat b^\dag\hat b^2\hat b^{\dag2}|0\rangle_b$. Since $\Omega\sim\omega\ll\omega_0$, we can safely approximate $\hat\Sigma_{1,2}(\Omega)$ by $\hat\Sigma_{1,2}(0)$ to obtain the effective Hamiltonian
\begin{equation}
\hat H_a=\hat H_{a0}+\hat\Sigma_1(0)+\hat\Sigma_2(0)=\omega\hat a^\dag\hat a-\frac{\lambda^2}{\omega_0}(\hat a^\dag+\hat a)^2+\frac{\lambda^4}{\omega^3_0N}(\hat a^\dag+\hat a)^4,
\end{equation}
which coincides with the unitary part in Eq.~(\ref{DFLG}).

\subsection{Mean-field analysis of the Lindblad-Landau theory}
At the mean-field level, Eq.~(\ref{GFLL}) implies the following equation of motion of $\alpha=\langle\hat a\rangle$:
\begin{equation}
i\partial_t\alpha=\left(\omega-i\frac{\kappa}{2}\right)\alpha-\Omega_2{\rm Re}\;\alpha+\frac{\Omega_4}{N}({\rm Re}\;\alpha)^3,
\end{equation}
which can be rewritten as
\begin{equation}
i\partial_t\tilde\alpha=\left(\omega-i\frac{\kappa}{2}\right)\tilde\alpha-\Omega_2{\rm Re}\;\tilde\alpha+\Omega_4({\rm Re}\;\tilde\alpha)^3
\label{rssemi}
\end{equation}
after the rescaling $\tilde\alpha\equiv\alpha/\sqrt{N}$. In addition to $\tilde\alpha_0=0$, when $\Omega_2>\Omega_{\rm c}=\omega+\frac{\kappa^2}{4\omega}$, Eq.~(\ref{rssemi}) has two fixed points 
\begin{equation}
\tilde\alpha_0=\pm\left(1+\frac{i\kappa}{2\omega}\right)\sqrt{\frac{\Omega_2-\Omega_{\rm c}}{\Omega_4}},
\label{fixpt}
\end{equation}
near which the semiclassical equation of motion (\ref{rssemi}) can be linearized to be
\begin{equation}
i\partial_t\delta\tilde\alpha=\left(\omega-i\frac{\kappa}{2}\right)\delta\tilde\alpha+[3\Omega_4({\rm Re}\;\tilde\alpha_0)^2-\Omega_2]{\rm Re}\;\delta\tilde\alpha.
\label{lnzsemi}
\end{equation}
The two eigenvalues of Eq.~(\ref{lnzsemi}) read
\begin{equation}
\lambda_\pm=-\frac{1}{2}[\kappa\pm\sqrt{\kappa^2-8\omega(\Omega_2-\Omega_{\rm c})}],
\end{equation}
which are both negative when $\Omega_2>\Omega_{\rm c}$, implying the stability of the two fixed points (\ref{fixpt}). After replacing $\lambda$ with $\lambda_t=\lambda_{t+T}$, we may expect that the lifetime of the damping mode besides the DTC mode is $O(\kappa^{-1})$. After a similar linearization-based analysis, we can obtain the two eigenfrequencies of the damping modes near $\tilde\alpha=0$ to be
\begin{equation}
\lambda'_\pm=-\frac{1}{2}[\kappa\pm\sqrt{\kappa^2+4\omega(\Omega_2-\Omega_{\rm c})}].
\end{equation}
As expected, both of $\lambda'_\pm$ are negative ($\lambda'_-$ becomes positive) when $\Omega_2<\Omega_{\rm c}$ ($\Omega_2>\Omega_{\rm c}$), implying the stability (instability) of the fixed point $\tilde\alpha=0$.

We note that Eq.~(\ref{fixpt}) can be used to perform a self-consistent check to justify the adiabatic elimination. Substituting $\Omega_2=\frac{4\lambda^2}{\omega_0}$ and $\Omega_4=\frac{2\Omega^2_2}{\omega_0}$ into Eq.~(\ref{fixpt}), we obtain
\begin{equation}
|\tilde\alpha_0|^2=\frac{\omega_0}{2\omega}\mu(1-\mu),
\label{dickealpha0}
\end{equation}
where $\mu=\frac{\lambda^2_{\rm c}}{\lambda^2}$ with $\lambda_{\rm c}=\frac{1}{2}\sqrt{\frac{\omega_0}{\omega}\left(\omega^2+\frac{\kappa^2}{4}\right)}$. While this result (\ref{dickealpha0}) differs generally from the order parameter $|\tilde\alpha|=\frac{\omega_0}{4\omega}(\mu^{-1}-\mu)$ in the original Dicke model, they do coincide near $\lambda=\lambda_{\rm c}$ or $\mu=1$. This provides an evidence that the effective theory does give a good approximation of the Dicke model in certain limits.

\subsection{Numerical calculation by exact diagonalization}
While the mean-field approximation should become exact in the large-$N$ limit, it is not clear how the lifetime of the DTC mode scales with respect to $N$. To convincingly show the exponentially long lifetime of the DTC mode, we return to the original Lindblad equation
\begin{equation}
\dot{\hat\rho}_t=-i(\hat H_{\rm eff}\hat\rho_t-\hat\rho_t\hat H^\dag_{\rm eff})+\kappa\hat a\hat\rho_t\hat a^\dag,
\label{effKerr}
\end{equation}
where the non-Hermitian effective Hamiltonian is given by
\begin{equation}
\hat H_{\rm eff}=\left(\omega-\frac{\Omega_2}{2}-i\frac{\kappa}{2}\right)\hat n+\frac{3\Omega_4}{16N}\hat n(\hat n+1)+\left(\frac{3\Omega_4}{16N}-\frac{\Omega_2}{4}\right)(\hat a^{\dag2}+\hat a^2)+\frac{\Omega_4}{8N}(\hat a^{\dag2}\hat n+\hat n\hat a^2)+\frac{\Omega_4}{32N}(\hat a^{\dag4}+\hat a^4).
\end{equation}
To employ exact diagonalization, we have to truncate the Hilbert space up to a $|n_{\rm max}\rangle$. While we expect that the profile of the Floquet steady state (approximately a Poisson distribution in the Fock space) is reliable as long as $n_{\rm max}\gtrsim2|\alpha_0|^2$, it is not clear whether the exponentially long life time could be reliable. Nevertheless, we can certificate the precision by changing $n_{\rm max}$ in practical numerical calculations.

In practice, we can perform exact diagonalization independently for the odd and even parity sectors, since Eq.~(\ref{effKerr}) respects the parity symmetry. To be concrete, if we choose the basis to be $|n\rangle\langle m|$ with $|m\rangle$ or $|n\rangle$ being a photon Fock state, then the Lindbladian in Eq.~(\ref{effKerr}) never mixes the sector with odd $m+n$ (odd parity) with that with even $m+n$ (even parity). This is true also for the Floquet-Lindblad superoperator. In particular, the steady state (DTC mode) can be found by diagonalizing the even-parity (odd-parity) sector. Using this approach, we obtain the numerical results presented in Fig.~\ref{fig4} in the main text.

\subsection{Possible phenomenology for asymmetric DTC behavior}
\begin{figure}
\begin{center}
        \includegraphics[width=17cm, clip]{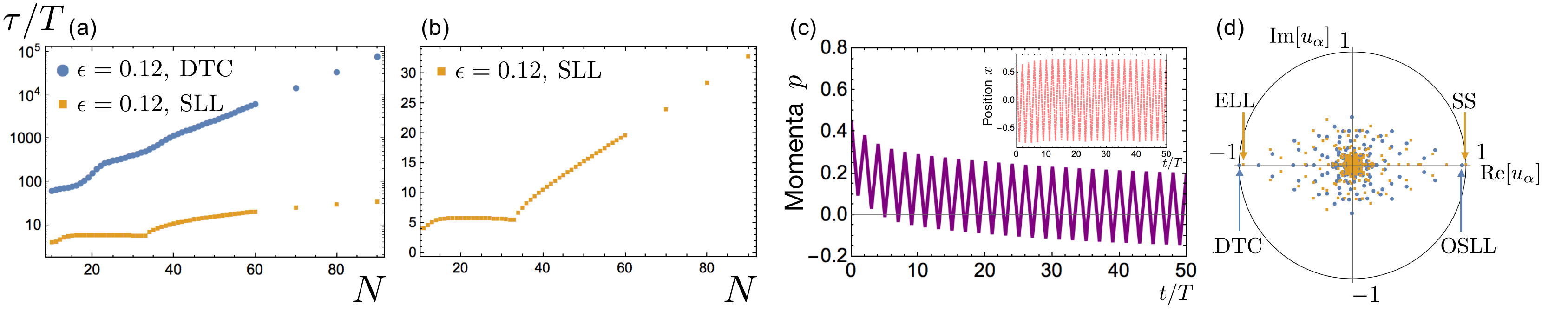}
      \end{center}
   \caption{(a) Finite-size scaling for the life time of the DTC mode and the second longest-lived mode in the Floquet-Lindblad-Landau model with the same protocol as in the main text except for $\epsilon=0.12$. (b) The same as (a) but the vertical axis is in the normal scale rather than the logarithmic scale. (c) Stroboscopic dynamics of rescaled $p=i\frac{\langle\hat a^\dag-\hat a\rangle}{\sqrt{2N}}$ and $x=\frac{\langle\hat a^\dag+\hat a\rangle}{\sqrt{2N}}$ (inset) starting form a coherent state (\ref{inistate}). The parameters are the same as (a) but for a fixed $N=80$ and the photon truncation is $n_{\rm max}=81$. (d) Floquet-Lindblad spectrum of the system in (c). The eigenvalues arising from the odd (even) sector are marked in blue (orange).}
      \label{figADTC}
\end{figure}

We have performed yet another finite-size scaling analysis for a large imperfection $\epsilon=0.12$, the semiclassical dynamics of which exhibits an asymmetric DTC behavior. As shown in Figs.~\ref{figADTC} (a) and (b), in addition to the exponentially long-lived DTC mode, the second longest-lived mode turns out to possess a relatively long life time that scales linearly with respect to $N$. In Fig.~\ref{figADTC} (c) we also present the stroboscopic dynamics of the rescaled quadratures starting from a coherent state
\begin{equation}
\hat\rho_0=|\alpha\rangle\langle\alpha|,\;\;\;\;\;\;\;\;|\alpha\rangle=e^{-\frac{|\alpha|^2}{2}}\sum_n\frac{\alpha^n}{\sqrt{n!}}|n\rangle,
\label{inistate}
\end{equation} 
with $\alpha=\sqrt{\frac{N}{5}}+\sqrt{\frac{N}{10}}i$. The dynamics turns out to be the relaxation of an asymmetric DTC (ADTC) behavior to the usual symmetric DTC order. Since the life time of the transient ADTC behavior seems to be consistent with that of the second longest-lived mode, it is natural to expect the latter to give rise to the former. In the thermodynamic limit, the second longest-lived mode persists and so does the ADTC behavior.

\begin{table*}[tbp]
\caption{Properties of several important Floquet-Lindblad eigenmodes, including the steady state (SS), the discrete time-crystalline (DTC) mode, the second longest-lived mode in the odd-parity sector (OSLL), and the longest-lived mode in the even-parity sector (ELL) . We use "$\sim$" to indicate the eventual decay of the mode after a long time.}
\begin{center}
\begin{tabular}{ccccc}
\hline\hline
\;\;State\;\; & \;\;Parity\;\; & \;\;Floquet-Lindblad eigenvalue\;\; & \;\;Odd-parity observable\;\; & \;\;Even-parity observable\;\; \\
\hline
SS & $+$ & 1 & 0 & const.\\
DTC & $-$ & $-1+O(e^{-cN})$ & $\sim+-+-\cdots$ & 0 \\
OSLL & $-$ & $1-O(N^{-1})$ & $\sim$const. & 0 \\
ELL & $+$ & $-1+O(N^{-1})$ & 0 & $\sim+-+-\cdots$ \\
\hline\hline
\end{tabular}
\end{center}
\label{table1}
\end{table*}

To establish a possible phenomenology for the ADTC behavior, it is constructive to look at the full Floquet-Lindblad spectrum (see Fig.~\ref{figADTC} (d)). After an intermediately large number of periods, e.g., $n\sim O(\sqrt{N})$, we can well approximate the state of the system by 
\begin{equation}
|\rho_{nT})\simeq|\rho_{\rm ss})+c_{\rm D}|\sigma_{\rm D})+c_{\rm O}|\sigma_{\rm O})+c_{\rm E}|\sigma_{\rm E}),
\label{rhoADTC}
\end{equation}
where $|\rho_{\rm ss})$, $|\sigma_{\rm D})$, $|\sigma_{\rm O})$ and $|\sigma_{\rm E})$ are the steady state, the DTC mode, the second longest-lived mode in the odd-parity sector parity and the longest-lived mode in the even-parity sector. As indicated by Fig.~\ref{figADTC} (d), the eigenvalues of $|\sigma_{\rm O})$ and $|\sigma_{\rm E})$ are almost symmetric with respect to the imaginary axis, implying that their life times are nearly the same. Further information of these four eigenmodes is summarized in Table~\ref{table1}. After a single period, the state of the system (\ref{rhoADTC}) evolves into
\begin{equation}
|\rho_{(n+1)T})\simeq|\rho_{\rm ss})-c_{\rm D}|\sigma_{\rm D})+c_{\rm O}|\sigma_{\rm O})-c_{\rm E}|\sigma_{\rm E}).
\label{rhoADTC2}
\end{equation}
Note that the coefficient of $|\sigma_{\rm O})$ does not flip the sign and thus gives rise to a basis for the DTC order resulting from $|\sigma_{\rm D})$ when we look at an odd-parity observable. Furthermore, we can infer from Eqs.~(\ref{rhoADTC}) and (\ref{rhoADTC2}) that the ADTC behavior emerges also in an even-parity observable, lasting for a time of the order of $O(N)$ before eventually relaxing to a constant instead of an exponentially long symmetric DTC order. In fact, this expectation has already been vindicated in the stroboscopic dynamics of $j_z$ in the top right panel in Fig.~\ref{figS1} as well as that in the rightmost panel in Fig.~\ref{fig2} in the main text.

\section{Details of the experimental implementations}
We here discuss concrete experimental implementation of the modulated open Dicke model in a cavity QED setup and its variation in a circuit QED setup. The main ideas are schematically illustrated in Fig.~\ref{figS5}.

\subsection{Cavity QED setup based on four-level atoms}
It is known that, as a result of the Thomas-Reiche-Kuhn (TRK) sum rule, the Dicke phase transition is always killed by the $A^2$ term in an equilibrium cavity QED setup \cite{Zakowicz1975}. However, the influence of the $A^2$ term can be neglected in the rotating frame in an intrinsically nonequilibrium setup based on the Raman transition, as pointed out in Ref.~\cite{Carmichael2007}. Such a proposal was first realized by using the atomic motional degrees of freedom \cite{Esslinger2010}, which correspond to a fixed $\omega_0\sim2\pi\times10\;{\rm kHz}$. This is not suitable for our proposal, since $\omega_0\ll\omega\sim2\pi\times10\;{\rm MHz}$ and the parity operator cannot be generated even approximately \cite{BUT}. Instead, we suggest that the experiment reported in Ref.~\cite{Parkins2014}, which is based fully on atomic internal states, might be an appropriate implementation of our proposal, where $\omega_0$, $\omega$ and $\lambda$ are of the same order of magnitude ($\sim2\pi\times1\;{\rm MHz}$). 

\begin{figure}
\begin{center}
\includegraphics[width=15cm]{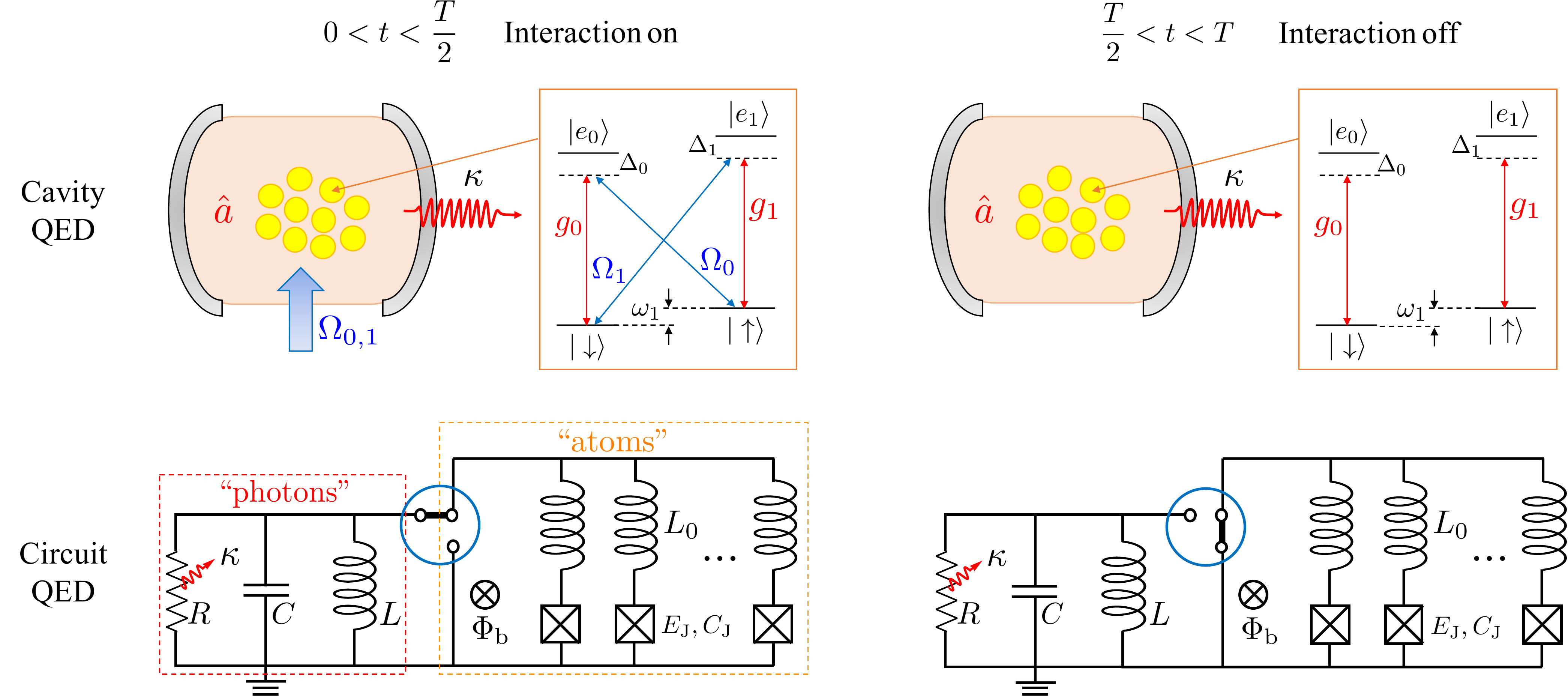}
\caption{Detailed implementations of the modulated Dicke model and its variation in the cavity (upper half) and circuit (lower half) QED systems. In the former case, the light-atom coupling is a Raman process assisted by excited states of four-level atoms. The coupling can be switched off if one stops shining the external driving lasers $\Omega_{0,1}$. In the latter case, the light-atom coupling is simulated by inductive coupling between an $RLC$ circuit (analogy of microwave photons with loss) and an array of Josephson oscillators (artificial atoms). The coupling can be turned off through a three-way switch (marked by blue cycles).}
\label{figS5}
\end{center}
\end{figure}

We first summarize the main results in Ref.~\cite{Carmichael2007}. Consider an ensemble of four-level atoms in an optical cavity with frequency $\omega_{\rm c}$. The four levels consist of two ground states $\ket{\downarrow}$, $\ket{\uparrow}$ and two excited states $\ket{e_0}$, $\ket{e_1}$, whose frequencies are $0$, $\omega_1$, $\omega_{\rm a0}$ and $\omega_{\rm a1}$, respectively. As shown in the upper half in Fig.~\ref{figS5}, the cavity mode interacts with the atom via the dipole transitions $\ket{\downarrow}\leftrightarrow\ket{e_0}$ and $\ket{\uparrow}\leftrightarrow\ket{e_1}$ with single-photon Rabi frequencies $g_0$ and $g_1$. Two additional classical driving lasers $(\omega_{\rm L0},\Omega_0)$ and $(\omega_{\rm L1},\Omega_1)$ are applied to couple $\ket{\uparrow}\leftrightarrow\ket{e_0}$ and $\ket{\downarrow}\leftrightarrow\ket{e_1}$, respectively. The frequencies satisfy $\omega_{\rm L1}-\omega_{\rm L0}\simeq2\omega_1$ and $\omega_{\rm L1}+\omega_{\rm L0}\simeq2\omega_{\rm c}$ in order to dramatically reduce the effective $\omega$ and $\omega_0$ in an appropriately chosen rotating frame. The detunings $\Delta_0\equiv\omega_{\rm a0}-\frac{1}{2}(\omega_{\rm L0}+\omega_{\rm L1})$ and $\Delta_1\equiv\omega_{\rm a1}-\omega_{\rm L1}$ are assumed to be so large that the excited-state manifold can be adiabatically eliminated. In this case, the effective Hamiltonian reads
\begin{equation}
\hat H_{\rm eff}=\hbar\omega\hat a^\dag\hat a+\hbar\omega_0\hat J_z+\hbar\delta\hat a^\dag\hat a\hat J_z+\frac{1}{\sqrt{N}}\hbar\lambda(\hat a^\dag\hat J_-+\hat a\hat J_+)+\frac{1}{\sqrt{N}}\hbar\lambda'(\hat a\hat J_-+\hat a^\dag\hat J_+),
\label{ADM}
\end{equation}
where the parameters are given by
\begin{equation}
\begin{split}
\omega=\omega_{\rm c}-\frac{1}{2}(\omega_{\rm L1}+\omega_{\rm L0})+\frac{N}{2}\left(\frac{g^2_0}{\Delta_0}+\frac{g^2_1}{\Delta_1}\right),&\;\;\;\;
\omega_0=\omega_1-\frac{1}{2}(\omega_{\rm L1}-\omega_{\rm L0})+\frac{1}{4}\left(\frac{\Omega^2_0}{\Delta_0}-\frac{\Omega^2_1}{\Delta_1}\right),\\
\delta=\frac{g^2_0}{\Delta_0}-\frac{g^2_1}{\Delta_1},\;\;\;\;
\lambda=&\frac{\sqrt{N}g_0\Omega_0}{2\Delta_0},\;\;\;\;
\lambda'=\frac{\sqrt{N}g_1\Omega_1}{2\Delta_1}.
\end{split}
\end{equation}
If we take an $A^2$ term $D(\hat a+\hat a^\dag)^2$ into account, the only difference is a small shift in $\omega$ by $2D$, which plays no role since $\omega$ is tunable via changing $\omega_{\rm L1}$ and $\omega_{\rm L2}$. The Dicke Hamiltonian can be obtained from Eq.~(\ref{ADM}) by fine-tuning the parameters of the external driving lasers such that $\delta=0$ and $\lambda=\lambda'$. Note that $N\sim10^5$ is also a tunable quantity. We therefore have enough degrees of freedom to independently control all the three parameters $\omega$, $\omega_0$ and $\lambda$ for the same cavity (with fixed $g_0$, $g_1$, $\omega_{\rm c}$ and $\kappa$).

To switch off the interaction in such a setup, we only have to switch off the driving lasers, corresponding to $\Omega_0=\Omega_1=0$ and thus $\lambda=\lambda'=0$. Note that $\omega$ and $\omega_1$ stay unchanged, since $\omega$ is independent of $\Omega_{0,1}$ and $\omega_0=\omega_1-\frac{1}{2}(\omega_{\rm L1}-\omega_{\rm L2})$, provided that $\delta=0$ and $\lambda=\lambda'$ are satisfied. Therefore, by simply switching on and off the external driving lasers as shown in the upper half in Fig.~\ref{figS5}, we can realize the modulated open Dicke model.

\subsection{Circuit QED setup based on inductive coupling}
As for the circuit QED setup based on superconducting qubits, we note that, due to the absence of the TRK sum rule for the capacitive coupling, the $A^2$ term could be negligible in the strong-coupling regime without entering the rotating frame of reference \cite{Ciuti2010} (although still controversial \cite{Marquardt2011,Rabl2016}). For the inductive coupling, while the Dicke phase transition has not yet been experimentally observed in superconducting circuits, the beyond-ultrastrong coupling has recently been realized for a single flux qubit \cite{Semba2017}. In a similar setup, the transient DTC order might be observable by fine-tuning the parameters and scaling up the number of superconducting qubits.

To be concrete, we discuss how to simulate a variation of the modulated open Dicke model by slightly modifying a circuit proposed in Ref.~\cite{Nakamura2016}, which has been demonstrated to exhibit a superradiant phase transition. As shown in the lower half in Fig.~\ref{figS5}, an $RLC$ circuit, which corresponds to a lossy ``photon" mode with frequency $\omega=(LC)^{-\frac{1}{2}}$ and decay rate $\kappa=(RC)^{-1}$, is integrated with an array of $N$ Josephson oscillators sharing the same flux bias $\Phi_{\rm b}=\frac{1}{2}\Phi_0$ ($\Phi_0\equiv\frac{h}{2e}$ is the flux quantum). The ``light-atom" coupling can be turned on/off via a three-way switch (marked by blue cycles). If the photon-like and atom-like circuits are coupled, the Hamiltonian reads
\begin{equation}
\hat H_1=\frac{\hat Q^2}{2C}+\frac{\hat \Psi^2}{2L}+\sum^N_{j=1}\left[\frac{\hat q_j^2}{2C_{\rm J}}+\frac{(\hat\psi_j-\hat \Psi)^2}{2L_0}+E_{\rm J}\cos\frac{2\pi\hat\psi_j}{\Phi_0}\right],
\label{BINH}
\end{equation}
where the charge operator $\hat Q$ ($\hat q_j$) and the flux operator $\hat\Psi$ ($\hat\psi_j$) of the $RCL$ circuit (the $j$th artificial atom) satisfy $[\hat\Psi,\hat Q]=i$ ($[\hat\psi_j,\hat q_k]=i\delta_{jk}$), $C_{\rm J}$ is the capacity of the Josephson junction and $E_{\rm J}$ is the Josephson energy. The \emph{plus} sign before $E_{\rm J}$ in Eq.~(\ref{BINH}) is due to the global flux bias $\Phi_{\rm b}$. This is crucial to enable the superradiant transition \cite{Nakamura2016}, which we believe would create  a transient DTC order even for small $N$.

Note that even if the $A^2$ ($\hat\Psi^2$ from $(\hat\psi_j-\hat\Psi)^2$) term is included, $\hat H_1$ still has an exact parity symmetry, i.e., the invariance under $\hat Q\to-\hat Q$, $\hat\Psi\to-\hat\Psi$, $\hat q_j\to-\hat q_j$ and $\hat\psi_j\to-\hat\psi_j$. This symmetry is maintained for the Lindblad equation where the jump operator is linear in $\hat Q$ and $\hat\Psi$. The parity operator can again be approximated as a time evolution under the following noninteracting Hamiltonian:
\begin{equation}
\hat H_2=\frac{\hat Q^2}{2C}+\frac{\hat \Psi^2}{2L}+\sum^N_{j=1}\left[\frac{\hat q_j^2}{2C_{\rm J}}+\frac{\hat\psi_j^2}{2L_0}+E_{\rm J}\cos\frac{2\pi\hat\psi_j}{\Phi_0}\right],
\end{equation}
provided the anharmonicity is small so that $E_{\rm J}\cos\frac{2\pi\hat\psi_j}{\Phi_0}\simeq E_{\rm J}-\frac{\hat\psi_j^2}{2L_{\rm J}}$ with $\hat L_{\rm J}=\frac{1}{E_{\rm J}}(\frac{\Phi_0}{2\pi})^2$, and the parameters satisfy $LC\simeq(\frac{1}{L_0}-\frac{1}{L_{\rm J}})^{-1}C_{\rm J}$. That is, the detuning between the $RLC$ circuit and the Josephson oscillators is small.

\end{document}